%% file: ChenLiTong_TPS_R1_v1a.tex
\documentclass[journal]{IEEEtran}
\usepackage{amsmath,amssymb,amsfonts,mathrsfs,mathtools,bm, bbm, dsfont,mathrsfs,amsthm,blkarray}
\usepackage[dvipdfm]{graphicx}
\usepackage[dvipsnames]{xcolor}

\usepackage{epsfig,epsf,psfrag,latexsym, graphics}
\usepackage[ruled,vlined]{algorithm2e}
\include{pythonlisting}

\usepackage{booktabs}
\usepackage{multirow,multicol}

\usepackage[breaklinks,colorlinks, linkcolor=MidnightBlue, anchorcolor=MidnightBlue, citecolor=MidnightBlue, urlcolor=MidnightBlue]{hyperref}
\usepackage[hyphenbreaks]{breakurl}
\usepackage{cite}
\usepackage{xurl}

\newcommand*{\QED}{\hfill\ensuremath{\square}}

 \def\old#1{}    

\usepackage{orcidlink}

\input LTmacros

\input{header.tex}

\begin{document}

\title{Multi-Interval Energy-Reserve Co-Optimization with SoC-Dependent Bids from Battery Storage}
\author{Cong Chen\orcidlink{0000-0002-0547-2889},~\IEEEmembership{Student Member,~IEEE,}
Siying Li\orcidlink{0000-0001-7118-7659},~\IEEEmembership{Student Member,~IEEE,}
and~Lang~Tong\orcidlink{0000-0003-3322-2681},~\IEEEmembership{Fellow,~IEEE}
\thanks{\scriptsize Part of the work is under review by the 2024 IEEE Power \& Energy Society General Meeting  (PESGM) \cite{LiChenTong22SoC}.}
\thanks{\scriptsize
Cong Chen, Siying Li, and Lang Tong (\url{{cc2662, sl2843 ,lt35}@cornell.edu}) are with Cornell University, Ithaca, NY 14853, USA. Cong Chen is also with Stanford University, Stanford, CA 94305. This work is supported in part by the National Science Foundation under Award 2218110 and 1932501, Power Systems and Engineering Research Center (PSERC) Research Project M-46, and the Stanford Energy Postdoctoral Fellowship.}}
\maketitle

\begin{abstract}
We consider the problem of co-optimized energy-reserve market clearing with state-of-charge (SoC) dependent bids from battery storage participants. While SoC-dependent bids capture storage's degradation and opportunity costs, such bids result in a non-convex optimization in the market clearing process. More challenging is the regulation reserve capacity clearing, where the SoC-dependent cost is uncertain as it depends on the unknown regulation trajectories ex-post of the market clearing. Addressing the nonconvexity and uncertainty in a multi-interval co-optimized real-time energy-reserve market, we introduce a simple restriction on the SoC-dependent bids along with a robust optimization formulation, transforming the non-convex market clearing under uncertainty into a standard convex piece-wise linear program and making it possible for large-scale storage integration. Under reasonable assumptions, we show that SoC-dependent bids yield higher profit for storage participants than that from SoC-independent bids. Numerical simulations demonstrate a 28\%-150\% profit increase of the proposed SoC-dependent bids compared with the SoC-independent counterpart.
\end{abstract}
   \vspace{-0.1in}

\section{Introduction} \label{sec:intro}
\input Intro_v5

\section{Co-Optimized Energy-Reserve Market Clearing with SoC-Dependent Bids} \label{sec:noncvxmodel}
\input ElectricityMarket_v5a

\section{Convexification of SoC-Dependent Energy-Reserve Market Clearing
} \label{sec:model}
\input CVXEM_v6a

\section{Profitability of  EDCR Bid } \label{sec:OptApprox}
\input OptEDCR_Bidnew_v6a


\section{Illustrative Example} \label{sec:Ex}
\input Example_v4

\section{Simulation} \label{sec:Sim}
\input Simulation_v6a

\vspace{-0.1cm}
 \section{Conclusion}\label{sec:conclusion}
This paper proposes a robust optimization for an energy-reserve co-optimized market to consider SoC-dependent bids from BESS. A subcategory of SoC-dependent bid, the EDCR bid, is characterized to reformulate the robust optimization into a convex piece-wise linear program, which is commonly used in the existing electricity market clearing process. Herein, the EDCR bid achieves the computation-effectiveness trade-off when considering SoC-dependent storage costs. Without bringing computationally expensive market clearing problems for the operator, the EDRC bid provides the flexibility and effectiveness to reveal the storage SoC-dependent costs compared with the SoC-independent storage bid in the current electricity market. Our analysis and simulation show that, under certain conditions, storage submitting the EDCR bid is used more frequently and receives higher profits compared with the SoC-independent bid.

\vspace{-0.1cm}

{
\bibliographystyle{IEEEtran}
\bibliography{BIB}
}

 
\newpage

\section{Appendix}
\input ExCapacityReg_v2
\input THM1proof_v5a
\input EDCRBidEX_v1
\input LPReform_v0
\input PropProof_v3a

\input Prop2Proof_v0
\input FinalProfitProof_v5a
\input simadd_v1
\end{document}

%% file: LTmacros.tex
\def\nn{\nonumber}
\def\beq{\begin{equation}}
\def\eeq{\end{equation}}
\def\bea{\begin{eqnarray}}
\def\eea{\end{eqnarray}}
\def\ba{\begin{array}}
\def\ea{\end{array}}

\def\bitem{\begin{itemize}}
\def\eitem{\end{itemize}}
\def\ben{\begin{enumerate}}
\def\een{\end{enumerate}}

\def\ie{{\it i.e.,\ \/}}

\def\alphabf{\hbox{\boldmath$\alpha$\unboldmath}}

\def\mubf{\hbox{\boldmath$\mu$\unboldmath}}

\def\upsilonbf{\hbox{\boldmath$\upsilon$\unboldmath}}
\def\phibf{\hbox{\boldmath$\phi$\unboldmath}}

\def\thetabf{{\mbox{\boldmath$\theta$\unboldmath}}}

\def\cbf{{\bm c}}
\def\dbf{{\bm d}}
\def\ebf{{\bm e}}

\def\gbf{{\bm g}}

\def\lbf{{\bm l}}
\def\mbf{{\bm m}}

\def\pbf{{\bm p}}
\def\qbf{{\bm q}}
\def\rbf{{\bm r}}
\def\sbf{{\bm s}}

\def\xbf{{\bm x}}
\def\ybf{{\bm y}}

\def\rbf{{\bm r}}
\def\xbf{{\bm x}}
\def\ybf{{\bm y}}

\def\Ebf{{\bm E}}

\def\Rbf{{\bm R}}
\def\Sbf{{\bm S}}

\def\Ec{{\cal E}}

%% file: header.tex
\newcommand{\beqa}{\begin{eqnarray}}
\newcommand{\eeqa}{\end{eqnarray}}
\newcommand{\beqan}{\begin{eqnarray*}}
\newcommand{\eeqan}{\end{eqnarray*}}

\newcounter{l1}
\newcounter{l2}
\newcounter{l3}
\newcommand{\bdotlist}{\begin{list}{$\bullet$}{}}
\newcommand{\bboxlist}{\begin{list}{$\Box$}{}}
\newcommand{\bbboxlist}{\begin{list}{\raisebox{.005in}{{\tiny
$\blacksquare$ \ \ }}}{}}
\newcommand{\bdashlist}{\begin{list}{$-$}{} }
\newcommand{\blist}{\begin{list}{}{} }
\newcommand{\barablist}{\begin{list}{\arabic{l1}}{\usecounter{l1}}}
\newcommand{\balphlist}{\begin{list}{(\alph{l2})}{\usecounter{l2}}}
\newcommand{\bAlphlist}{\begin{list}{\Alph{l2}.}{\usecounter{l2}}}
\newcommand{\bdiamlist}{\begin{list}{$\diamond$}{}}
\newcommand{\bromalist}{\begin{list}{(\roman{l3})}{\usecounter{l3}}}

\newtheorem{theorem}{Theorem}
\newtheorem{lemma}{Lemma}
\newtheorem{proposition}{Proposition}

\newtheorem{definition}{Definition}

%% file: Intro_v5.tex
Battery energy storage systems (BESS) are exceptional resources in transmission systems with substantial renewable integration. Large-scale participation of BESS can shave peak load, reduce the intermittency of renewable energy, and provide fast-ramping and high-quality regulation services. However, in a deregulated bid-based electricity market, a merchant BESS participant faces the challenge of constructing bids and offers that accurately capture its costs. Unlike traditional generation resources, the costs of BESS participants include battery degradation and opportunity costs, both functions of the scheduled state-of-charge (SoC) trajectory \cite{CAISO_DEB:20, Ecker14JPSdegradaES}.   In particular, the health of BESS is closely related to the battery charge cycle, and the battery SoC at one time directly affects its ability to capture future profitable charging/discharging opportunities. Therefore, there is valid interest in considering SoC-dependent bidding, allowing BESS participants to express their willingness to charge/discharge as a function of battery SoC \cite{CAISO_SOCdependent:22, ZhengXu22energy}.   To this end, this work aims to shed light on the {\em computation-effectiveness trade-off} when SoC-dependent bids are allowed in the energy and reserve markets: (i) For the market operator, can the existing market clearing framework accommodate {\em computationally} expensive SoC-dependent bids? (ii) For BESS, do SoC-dependent bids {\em effectively} reveal storage cost and bring meaningful profit gain over SoC-independent bids? 

In this work, we consider SoC-dependent bids from BESS participants in a co-optimized energy-reserve real-time market, where BESS participants bid energy charging/discharging services and regulation reserve capacity.\footnote{For simplicity, we only consider regulation reserve and energy markets, which can be extended to other types of reserve markets in \cite{PJM}.}  In particular, this paper considers a multi-interval look-ahead dispatch of BESS and other generation resources, where BESS are scheduled for a block of future intervals with only the immediate (binding) interval implemented. We balance the BESS bidding model effectiveness and computation efficiency in the market clearing problem, which uses the multi-interval look-ahead dispatch to determine energy charging/discharging dispatch quantities and reserve capacities. Through analysis and empirical studies, we evaluate BESS's profitability and power system operation costs of having SoC-dependent bids.


   \vspace{-0.1in}
\subsection{Summary of contribution}

We address the nonconvexity and uncertainty in a multi-interval co-optimized energy-reserve market involving SoC-dependent bids. Our contribution is threefold.

First, we present a robust formulation of the co-optimized energy-reserve market with SoC-dependent bids. The regulation reserve capacity is cleared when the SoC-dependent storage cost is uncertain, as it depends on the unknown regulation trajectories ex-post of the market clearing. The novelty of our formulation is a rigorous model of such uncertainties in the multi-interval look-ahead energy and regulation dispatch, compared with the ad-hoc characterization of BESS costs in the existing electricity market.

Second, we show in Theorem~\ref{thm:Eq} that the nonconvex energy-reserve
co-optimization can be convexified if the BESS bids satisfy the so-called Equal Decremental-Cost Ratio (EDCR) condition \cite{ChenTong22arXivSoC}, rendering the intractable robust energy-reserve co-optimization a standard convex piecewise linear program commonly used in the existing market clearing program. Such convexification is crucial for the large-scale integration of BESS. The convexification is also significant when pricing energy and reserve services by simplifying the problem of uplift payments \cite{Cong&Tong:22Allerton}.

Third, we establish a performance guarantee and numerical evaluation of SoC-dependent EDCR bids over standard SoC-independent bids. In particular, we show in Theorem~\ref{thm:ProfitEnergy}  that, given any SoC-independent
bids, we propose an optimization to find  EDCR bids with higher storage profitability under certain assumptions. We further relax these assumptions in the simulation and empirically demonstrate that EDCR bids can increase storage profits by 28\%-150\% over SoC-independent bids in the energy-reserve co-optimized market.

\vspace{-0.05in}
\subsection{Related work}


BESS, especially lithium battery, are known to have SoC-dependent costs, including the lifetime degradation cost \cite{ Ecker14JPSdegradaES, shi18TACrainflow, MaMei24TEMPR, Chowdhury&etal:24JES} and opportunity cost \cite{CAISO_DEB:20, JiangPowell15Storage, Lamadrid24JRE, Zhou&etal:24ACM}. When operated at different SoC levels, BESS have different degradation rates and different capabilities in catching future arbitrage profit opportunities.\footnote{For example, BESS have opportunity costs when charging at low prices before capturing future arbitrage profits to discharge at high prices.} Literature considering these costs generally falls into two categories. Type I starts from the perspective of BESS to quantify these costs and produce bidding strategies in the electricity market. Type II starts from the perspective of the power system and market operator to design dispatch and market-clearing models incorporating these BESS costs.

Extensive research within Type I focuses on SoC-independent bidding methods, such as quantity bids \cite{shi18TACrainflow, HarshaDahleh15TPS} and price-quantity bids \cite{JiangPowell15Storage, Lamadrid24JRE}. They are grounded in the fact that current electricity markets permit BESS to self-dispatch or submit bid-in prices revealing both its willingness to charge and discharge, independent of SoC \cite{CAISO_StorageBid:21, ERCOT_storage:24}. Existing studies typically aim to minimize storage degradation costs, maximize storage arbitrage opportunity profits, and focus on the energy market with volatile prices. Some emerging research explores the joint energy-regulation market considering uncertainties of AGC signals \cite{MaMei24TEMPR}, and SoC-dependent bidding methods using dynamic programs \cite{ZhengXu22socAribitrage, Zhou&etal:24ACM}.

Previous research in Type II typically incorporates BESS degradation costs by the mixed integer program (MIP) \cite{ ZhaoQiu18robustESloss, Vykhodtsev24TSE}. Recent Type II research has run in parallel with the industry's interest, the proposal by California Independent System Operator (CAISO) to allow new SoC-dependent bids, capturing various BESS costs including opportunity and degradation costs\cite{CAISO_SOCdependent:22}. However, such  SoC-dependent bids result in a computationally expensive non-convex market clearing problem that necessitates the use of MIP \cite{ZhengXu22energy}. The nonconvexity of the market clearing problem brought by the SoC-dependent bid also presents pricing challenges, resulting in out-of-merit dispatch and the need for out-of-the-market settlements in CAISO's real-time market. Additionally,  a new bidding format for degradation costs proposed by \cite{bansal22EPSRrainflowmarket} is cleared by a convex energy market. However, this format cannot capture opportunity costs and limits the bid-in price to BESS cycle depths, restricting the storage's ability to express its willingness to buy and sell energy across different SoC levels.

The first half of our research falls under Type II, focusing on reducing the {\em computation} burden for the market operator when incorporating SoC-dependent bids in the energy-reserve joint market. Our convexification method, the EDCR condition, was first discovered in a short communication  \cite{ChenTong22arXivSoC}  under a more restricted setting that involves only multi-interval energy market clearing. We formalize and extend the results from \cite{ChenTong22arXivSoC, Cong&Tong:22Allerton} to the convexification of the energy-reserve co-optimized real-time market. The second half of our research aligns with Type I. To support the {\em effectiveness} of EDCR bids, we provide a theoretical quantification for the profitability of EDCR bids and a method to generate EDCR bids from SoC-independent bids with improved profits. Together, our findings show that EDCR bids achieve the {\em computation-effectiveness trade-off} when incorporating SoC-dependent storage costs in the energy-reserve joint market. 


   \vspace{-0.15in}
\subsection{Paper organization and notations}

Incorporating the SoC-dependent bid in the electricity market involves both the market operator and BESS. We focus on the role of the operator in Sec.~\ref{sec:noncvxmodel}-Sec.~\ref{sec:model}, and the role of storage is explained in Sec.~\ref{sec:OptApprox}. In Sec.~\ref{sec:noncvxmodel}, we first introduce the storage participation model in the co-optimized energy-reserve market. Then, we establish a robust optimization, clearing SoC-dependent bids in the energy-reserve market, which is intractable because of the nonconvexity when computing the SoC-dependent cost of storage. In Sec.~\ref{sec:model}, we present the convexification condition, the EDCR condition, for this nontrivial robust optimization.  In Sec.~\ref{sec:OptApprox}, we proposed a method to produce EDCR  bids with profitability compared with the SoC-independent bid. Related toy examples and simulations are in Sec.~\ref{sec:Ex} and Sec.~\ref{sec:Sim}. 

For the notations, we use $x$ for scalar, $\xbf:=(x_i)$ for the vector of components $x_i$, ${\bf 1}$ for the vector of 1's, and define $[x]:=\{1,...,x\}$. The indicator function $\mathbbm{1}$ means that  $\mathbbm{1}_{\cal X}=1$ if   ${\cal X}$ is true, and $\mathbbm{1}_{\cal X}=0$ otherwise. The notation $\xbf \neq \ybf$ means $\exists i, s.t~~ x_i \neq  y_i$. All symbolic superscripts in this paper are used to distinguish notations rather than represent the power of a number. Major symbols are listed in Table~\ref{tab:symbols}.

{\small
\begin{table}[htbp]
\caption{Major symbols}\label{tab:symbols}
\begin{center}
\vspace{-1em}
\begin{tabular}{ll}
\hline
$T\in \mathbb{R}_+$& total number of $\tau$-minute (eg. $\tau=$ 15 ) intervals \\
$\pbf^c_i,\pbf^d_i \in \mathbb{R}^T$ & charging and discharging  power  of unit $i$\\
$K\in \mathbb{R}_+$& total number of segments for SoC-dependent bid\\
$\cbf^d_i,\cbf^c_i \in \mathbb{R}^K$ & bid-in  price for charge/discharge  power\\
$\rbf^u_i,\rbf^d_i\in \mathbb{R}^T$ &  regulation up and down capacity of unit $i$\\
$\qbf^c_i,\qbf^d_i\in \mathbb{R}^T$ & net input and net output of unit $i$\\
$t\in  [T]$& index of the $\tau$-minute time interval\\
$J\in \mathbb{R}_+$& total number of $\delta$-s (eg. $\delta=$ 4) in interval $t$\\
$\mbf^u_t,\mbf^d_t \in \mathbb{R}^J $ &  regulation up and down mileage in interval $t$\\
\hline
\end{tabular}
\vspace{-2em}
\end{center}
\end{table}
}

%% file: ElectricityMarket_v5a.tex
This section presents a multi-interval energy-reserve co-optimization for the market operator to incorporate the SoC-dependent bids from BESS, generalized from the standard real-time energy-reserve co-optimization \cite{PJM, CAISO_ESReg23}. A novel contribution here is a robust optimization in evaluating SoC-dependent costs associated with yet-realized regulation dispatch of storage. The convexification condition, discussed in the following section, enables the market operator to solve this robust optimization by a linear program, matching the existing electricity market clearing in real-time.
    \vspace{-0.15in}
\subsection{Storage operation model in energy-reserve co-optimization}
 We consider BESS participating in the energy-reserve co-optimization, where the energy dispatch and regulation reserve capacities are determined. The BESS regulation-reserve capacities are cleared in  real-time with granularity $\tau$ (eg. $\tau=$ 15 minutes in \cite{CAISO_RegulationMarket:22}) and the time index is $t$. For simplicity, we consider energy and reserve with the same time granularity $\tau$ here, and the model can be easily extended to  different time granularities. In the scheduling interval $t$, we denote $e_t$ for the storage SoC, $p_t^c$ for the charging power, $p_t^d$ for the discharging power, $r^d_t$ for the regulation down capacity, and $r^u_t$ for the regulation up capacity.  The storage SoC evolves by
\begin{subequations}\label{eq:SoCevolvePP}
\begin{align}
e_{t+1}=e_{t} +\eta&(p^c_{t}+\gamma^d_{t}r^d_{t})\tau-(p^d_{t} +\gamma^u_{t}r^u_{t} )\tau,\label{eq:SoCevolveE}\\ &p^c_{t}p^d_{t}=0,\label{eq:Simul_CD}
\end{align}
\end{subequations}
where $\eta\in (0,1]$ is the round trip efficiency of the storage.\footnote{Our model can be easily extended to include separate storage charging and discharging efficiencies. Relevant research is shown in \cite{ChenTong22arXivSoC}.} $\gamma^d_t$ and $\gamma^u_t$ are the expected values of the multipliers   indicating the fraction of actually utilized regulation down and up capacities\footnote{More explanations about $\gamma^d_t$ and $\gamma^u_t$ are shown in equation (2) of \cite{CAISO_ESReg23}.}.  As is constrained by \eqref{eq:Simul_CD},  in the energy market, storage is not scheduled to charge and discharge simultaneously. However, in the regulation market, storage can provide both regulation up and regulation down capacities  \cite{CAISO_StakeholderComments:22}. To guarantee that storage operations are bounded by its SoC capacity limits when bi-direction regulation capacities are cleared, we have  
\beq\label{eq:SoClimit}
\begin{aligned}
e_{t}+\eta(p^c_{t} +\gamma^d_{t}r^d_{t} )\tau   \leq \bar{E}, ~~
\underline{E}\le e_{t}-(p^d_{t} +\gamma^u_{t}r^u_{t} )\tau,
\end{aligned}
\eeq
where $\bar{E}$ and $\underline{E}$ represent the maximum and minimum SoC limits of BESS, respectively.

\vspace{-0.15in}
\subsection{SoC-dependent bids}\label{sec:SOCbid}

\begin{figure}[h]
\centering
\includegraphics[scale=0.38]{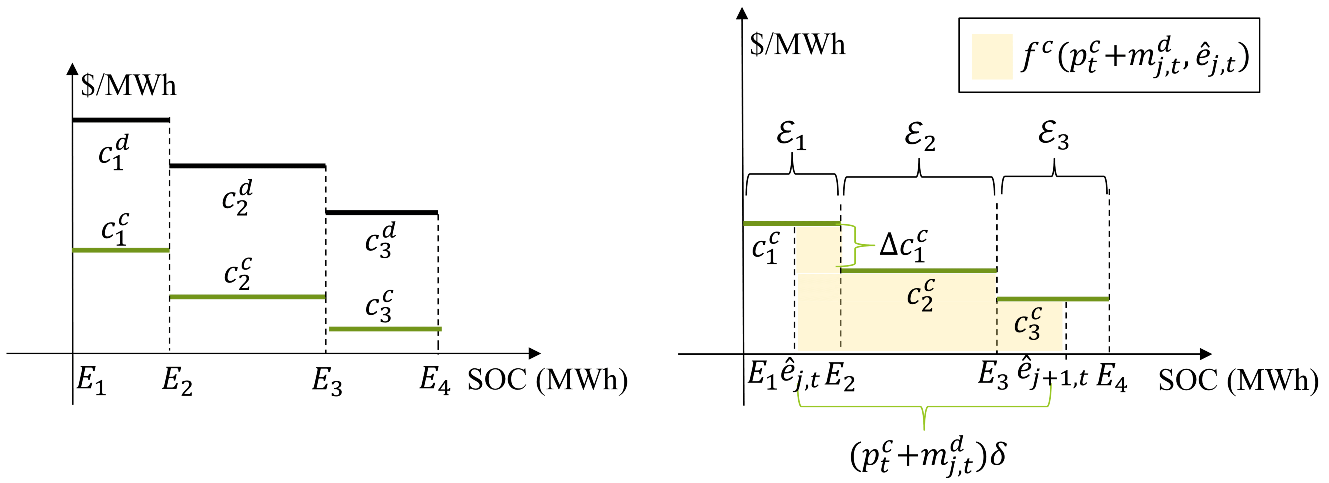}
\caption{\scriptsize  Left: The SoC-dependent bid and offer format when $K=3$. Right: Cost of charging the storage by $(p^c_{t}+m^d_{j,t})\delta $ from $\hat{e}_{j,t}$ to $\hat{e}_{j+1,t}$. }
\vspace{-0.25cm}
\label{fig:SOC_D_Bid}
\end{figure}

A standard piecewise-linear SoC-dependent bid model \cite{CAISO_SOCdependent:22} is illustrated in Fig.~\ref{fig:SOC_D_Bid} (left). Without loss of generality, we partition the SoC axis into $K$ consecutive segments, \ie
\beq\label{eq:Ekk}
\underline{E}\le E_{k}\le E_{k+1} \le \overline{E}, \forall k \in [K].
\eeq
Within each segment $\Ec_k:=[E_k, E_{k+1}]$, a pair of bid-in {\em marginal cost/benefit} parameters  $(c_k^c, c_k^d)$ is defined. Denote $\Ebf:=( E_{k}), \cbf^c:=(c^c_k)$, and $\cbf^d:=(c^d_k)$. The marginal discharging (bid-in) costs (to the grid) $\tilde{b}^d(e_t; \mathbf{c}^d, \mathbf{E})$ and marginal charging (bid-in)  benefits (from the grid) $\tilde{b}^c(e_t; \mathbf{c}^c,\mathbf{E})$ are functions of battery SoC $e_t$. In particular, using the indicator function\footnote{$\mathbbm{1}_{\{s \in \Ec_i\}}$ equals to 1 when $s \in \Ec_i$.}  $\mathbbm{1}$,  
\beq\label{eq:SoCBid}
\begin{array}{r}
\left\{\begin{array}{ll}
\tilde{b}^c(e_t; \mathbf{c}^c,\mathbf{E}):=\sum_{k=1}^Kc^c_k\mathbbm{1}_{\{e_t\in \Ec_k\}},\\
\tilde{b}^d(e_t; \mathbf{c}^d,\mathbf{E}):=\sum_{k=1}^Kc^d_k\mathbbm{1}_{\{e_t\in \Ec_k\}}.
\end{array}\right.
\end{array}
\eeq

For the longevity of BESS and the ability to capture profit opportunities, it is more costly to discharge when the SoC is low, and the benefit of charging is small when the SoC is high. Therefore, typical  bid-in discharging costs $(c^d_{k})$ and charging benefits $(c^c_{k})$ are monotonically decreasing. Furthermore,  the storage participant is willing to discharge only if the selling price is higher than the buying price. Hence, BESS's willingness to sell by discharge must be higher than its willingness to purchase (adjusted to the round trip efficiency),  \ie $c^d_{\mbox {\tiny K}}> c^c_{1}/\eta$.  Together,  SoC-dependent bids and offers satisfy the following.
\begin{definition}[Monotonic SoC-Dependent Bids] \label{assume:single} An SoC-dependent bids parameterized by $\{\cbf^c, \cbf^d, \eta\}$ is monotonic if 
\beq\label{eq:a1}
c_k^c \ge c_{k+1}^c,~
c_k^d \ge c^d_{k+1},~{\rm and}~ c^c_{1}/\eta < c^d_{\mbox {\tiny K}}, \forall k \in [K-1].\eeq
\end{definition}

Storage marginal costs, including battery degradation and opportunity costs, are generally nonlinear and computationally expensive if directly submitted to the electricity market. That way, the electricity market provides specific bidding formats, such as the SoC-independent\footnote{For SoC-independent bid, we set $c^c_k=\bar{c}^c, c^d_k=\bar{c}^d, \forall k \in [K]$.} or SoC-dependent bid, to reduce computation burdens and allow storage to reveal its {\em bid-in cost}. The method to compute storage bid-in costs from SoC-dependent bids is introduced in the following section.

   \vspace{-0.15in}
\subsection{Bid-in cost of co-optimized storage operation}


Storage reserves regulation capacities for real-time frequency regulation services and the actual storage operation is correlated with the regulation mileage, which follows the Automatic Generation Control (AGC) signal delivered with time granularity $\delta$ (eg. $\delta=$ 4-s \cite{Chen:15}). During the $t$-th time interval, we denote $m^u_{j,t}, m^d_{j,t}$ as the regulation-up  and regulation-down mileage, respectively. We have
\beq\label{eq:RegM}
\sum_{j=1}^J(m^u_{j,t}\delta)=\gamma^u_t r^u_t\tau,~~ 
\sum_{j=1}^J(m^d_{j,t}\delta)= \gamma^d_t r^d_t\tau,
\eeq
where $J$ denotes the number of $\delta$-s time intervals within interval $t$. Denote $\hat{e}_{j,t}$ as  SoC at the beginning of the $j$-th interval during time $t$, which evolves by
\begin{subequations}\label{eq:SoC4s}
\begin{align}
     &&& \hat{e}_{j+1,t} = \hat{e}_{j,t}+ \eta( p^c_{t} + m^d_{j,t} )\delta - ( p^d_{t}  + m^u_{j,t} )\delta,\\
    &&& m^u_{j,t} m^d_{j,t}=0, ~~e_t =\hat{e}_{1,t}, ~~e_{t+1} =\hat{e}_{J+1,t}. \label{eq:mSim}
\end{align}
\end{subequations}
As is constrained by \eqref{eq:mSim},  AGC won't require storage to regulate up and down  simultaneously \cite{ERCOTReg:23}, and the beginning and end state SoC with granularity $\delta$ equal to the SoC correspondents  at $t$ and $t+1$.



The SoC-dependent bid-in cost for storage during interval $t$  is given by
\beq \label{eq:SoCcostMile}
f(\mbf^u_t,\mbf^d_t, e_t):=  \sum_{j=1}^J f^d(p^d_{t} + m^u_{j,t}, \hat{e}_{j,t} )-f^c(p^c_{t} + m^d_{j,t}, \hat{e}_{j,t}), 
\eeq
which is storage discharging cost minus the charging benefits.

Denote $\Delta c^d_{k}:=c^d_{k}-c^d_{k+1}$,  $\Delta c^c_{k}:=c^c_{k}-c^c_{k+1}$, and $m, n$ as indexes for SoC segments with $ \hat{e}_{j,t} \in \Ec_m$ and   $\hat{e}_{j+1,t}\in \Ec_{n}$. The SoC-dependent discharging and charging costs are  
\begin{align}
    &f^d(x, y):=\mathbbm{1}_{\{n\le m\}}c^d_{n}\delta x+\mathbbm{1}_{\{n < m\}}\sum_{k=n+1}^{m} \Delta c^d_{k-1}(E_{k}- y),\nn\\
    &f^c(z, y):=\mathbbm{1}_{\{n\geq m\}}c^c_{n}\delta z+\mathbbm{1}_{\{n > m\}}\sum_{k=m}^{n-1}\frac{\Delta c^c_{k}}{\eta}(E_{k+1}- y),\nn
\end{align}
where $x$, $y$, and $z$ are used as symbols rather than variables with specific physical meanings. In Fig.~\ref{fig:SOC_D_Bid} (right), we provide an example of storage charging cost with SoC dependency. The SoC-dependent cost in  \eqref{eq:SoCcostMile} relies on the regulation mileage signals $\mbf^u_t,\mbf^d_t$ for accurate computation. This means, in the SoC-dependent electricity market, the bid-in cost depends on the SoC trajectory following AGC, \ie $\hat{\ebf}_t:=(\hat{e}_{1,t},...,\hat{e}_{j,t}...,\hat{e}_{J+1,t})$. In Appendix~\ref{sec:Ex_SOCdep}, we provide a toy example showing storage with the same regulation and energy capacities cleared has different SoC-dependent costs when the SoC trajectories $\hat{\ebf}_t$ are different. 

However, regulation signals are unknown when clearing regulation capacities in the real-time market. Therefore, we provide a robust optimization to clear the SoC-dependent energy-reserve market. Such a robust  optimization finds the most costly SoC trajectory when clearing SoC-dependent bids for energy and regulation capacities. 
 
 We obtain the worst-case SoC-dependent storage bid-in cost and  the worst-case SoC trajectory $\hat{\ebf}_t$ within time $t$ via
\begin{subequations} \label{eq:ROmileage}
    \begin{align}
f^*(p^c_t, p^d_t, r^u_t, r^d_t,e_t) := &\underset{  \hat{\ebf}\in [\underline{E},\overline{E}]^J }{\rm maximize} && f(\mbf^u_t,\mbf^d_t, e_t ), \\
&\mbox{subject to}&& \eqref{eq:RegM}, \eqref{eq:SoC4s}, \forall j \in [J].
\end{align}
\end{subequations} 
Here, the cleared quantities of energy dispatch (eg. $p^c_t, p^d_t$) and regulation-reserve capacities (eg. $r^u_t, r^d_t$) are given, and the objective takes the storage bid-in cost formula in \eqref{eq:SoCcostMile}.

Considering $T$-interval look ahead dispatch when clearing the joint energy-reserve market with time granularity $\tau$, the SoC-dependent bid-in storage cost for the $T$-interval is  
\beq\label{eq:robustcost}
F^*(\pbf^c, \pbf^d, \rbf^u, \rbf^d;s) :=\sum_{t=1}^T f^*(p^c_t, p^d_t, r^u_t, r^d_t,e_t),
\eeq
where $s$ is the initial SoC of storage, \ie $e_1=s$.

\vspace{-0.1in}
\subsection{Energy-reserve co-optimization as a robust optimization}

Considering the worst-case SoC-dependent cost \eqref{eq:robustcost} for storage, we propose the robust optimization \eqref{eq:NONCVX} to minimize the system operation cost for providing certain energy dispatch and regulation capacities. We introduce the storage index $i$ to include multiple BESS in the market clearing problem. For simplicity, we establish the dispatch model over a $M$-bus power network  with one generator and one storage at each bus, which is extendable to general cases. 

For interval $t$, let $d_{it}$ represent the inelastic demand at bus $i$, and $\dbf[t]:=(d_{1t},\cdots, d_{Mt})$ denote the demand vector for all buses. Similarly, $\gbf^e[t]$ is the generator dispatch in the energy market, and $\pbf^d[t]$ and $\pbf^c[t]$ are the vectors of charging and discharging power of BESS at different buses. $\xi_t^u$ and $\xi_t^d$ are zonal regulation requirements at time $t$. Define  $\Omega := \{\gbf_i^e, \gbf_i^u, \gbf_i^d,\pmb{p}_{i}^c,\pmb{p}_{i}^d, \pmb{r}_{i}^u,\pmb{r}_{i}^d \in \Rbf^T_+ \}$ for the domain. Given the initial SoC $e_{i1}=s_i$ and the load forecast $(\dbf[t])$ over the $T$-interval scheduling horizon, the economic dispatch that minimizes the system operation costs is as follows.
\begin{subequations}
\label{eq:NONCVX}
\begin{align}
& \underset{\substack{\Omega}}{\rm minimize} && \sum_{i=1}^{M} (h_i(\gbf^e_i, \gbf_i^{u}, \gbf_i^{d})+ F^*_{i}(\pbf^c_i, \pbf^d_i, \rbf^u_i, \rbf^d_i;s_i))\label{eq:obj} \\
& \mbox{subject to}&& \forall t\in [T], \forall i\in  [M],\nn \\ \label{eq:PFlimit}
&&& \pmb{S} (\pmb{g}^e[t]+\pbf^d[t]-{\pbf}^c[t]-\pmb{d}[t]) \le \pmb{l},\\ \label{eq:PF}
&&& {\bf 1}^\intercal(\pmb{g}^e[t]+\pbf^d[t]-{\pbf}^c[t]-\pmb{d}[t]) = 0,\\ \label{eq:ReqReg1}
&&& \sum_i^M (g_{it}^{u} + r_{it}^u) \geq \xi_t^u, \\ \label{eq:ReqReg2}
&&& \sum_i^M (g_{it}^{d} + r_{it}^d) \geq \xi_t^d, \\ \label{eq:gen}
&&& g^e_{it}+g^u_{it}\leq \bar{g}_i,~~ \underline{g}_i \le g^e_{it}-g^d_{it}, \\ \label{eq:SoCcons1}
&&& e_{i1}=s_i, ~~\eqref{eq:SoCevolvePP}, ~~\eqref{eq:SoClimit},\\ \label{eq:SoCcons2}
&&& 0\le r^d_{it}\le \bar{r}^d_i, ~~0\le r^u_{it}\le \bar{r}^u_i,\\ \label{eq:SoCcons3}
&&& 0 \le p^d_{it}\le \bar{p}^d_i, ~~0 \le p^c_{it}\le \bar{p}^c_i,
    \end{align}
\end{subequations}
where  $h_i(\gbf^e_i, \gbf_i^{u}, \gbf_i^{d})$ in \eqref{eq:obj} is the convex piecewise linear bid-in cost from the $i$-th generator, $\gbf^e_i$, $\gbf_i^{u}$, and $\gbf_i^{d}$ represent the power in the energy market, the regulation up capacity, and the regulation down capacity. The DC power flow model is considered in \eqref{eq:PFlimit}-\eqref{eq:PF} with the shift-factor matrix $\Sbf \in \mathbb{R}^{2B\times M}$ for a network with $M$ buses and $B$ branches (branch flow limit $ \lbf \in \mathbb{R}^{2B}$). Other system operation constraints include regulation capacity requirements \eqref{eq:ReqReg1}-\eqref{eq:ReqReg2}, generator capacity limits \eqref{eq:gen}, and storage constraints, comprising SoC state-transition constraints, SoC limits, and storage capacity limits \eqref{eq:SoCcons1}-\eqref{eq:SoCcons3}. 

Such a robust  optimization is intractable for two reasons. First, the SoC-dependent storage cost $F^*$ in the objective function relies on \eqref{eq:ROmileage}, which has a nonconvex objective function and bilinear constraints. Second, the equality constraint \eqref{eq:Simul_CD}, prohibiting simultaneous charging and discharging energy dispatch, is bilinear.  


To facilitate an efficient energy-reserve co-optimization that can be computed with the time granularity $\tau$, we propose a convexification condition in the following section to solve this intractable robust optimization.

%% file: CVXEM_v6a.tex
This section addresses the issue of nonconvexity of the energy-reserve co-optimization involving SoC-dependent bids. 

\vspace{-1em}
\subsection{EDCR bids and convexification}
We introduce the  Equal Decremental-Cost Ratio (EDCR) condition on the SoC-dependent bidding format with which the energy-reserve co-optimization \eqref{eq:NONCVX} becomes convex.\footnote{Storage index $i$ is omitted in Definition~\ref{def:EDCR} and Theorem~\ref{thm:Eq} for simplicity.}  
 \begin{definition}[EDCR Condition] \label{def:EDCR}
A $K$-segment monotonic SoC-dependent bid with parameters $\cbf^c, \cbf^d$, and $\eta$ satisfies the EDCR condition if 
\beq\label{eq:EDCR}
\frac{c^c_{k+1}-c^c_{k}}{c^d_{k+1}-c^d_{k}}=\eta, \forall k \in [K-1] .
\eeq
 \end{definition}
EDCR bid adds extra conditions for the monotonic SoC-dependent bids in Definition~\ref{assume:single}. Herein, SoC-dependent bids satisfying the EDCR condition are referred to as {\em EDCR bids}. EDCR bids can match storage opportunity cost in arbitrage, which is usually evaluated by storage profit maximization with price forecasts. In the example of Sec.~\ref{sec:EX_EDCR}, we derive the storage opportunity cost with the stochastic dynamic storage profit maximization\cite{ZhengXu22energy, Zhou&etal:24ACM} and find it an EDCR bid. The EDCR condition in equation \eqref{eq:EDCR} requires that the charging cost divided by the discharging cost equals storage round trip efficiency. This means the discharging power is more valuable than the charging power because of the storage efficiency loss in catching the opportunity cost. EDCR bid cannot fit general storage degradation costs. In these cases, we propose the EDCR bid generation method in Sec.~\ref{sec:OptApprox} to approximate the general storage cost with EDCR bid.

We now convexify the objective function and relax the bilinear equality constraints of the energy-reserve market clearing problem (\ref{eq:NONCVX}). Theorem~\ref{thm:Eq} below convexifies the worst-case storage cost \eqref{eq:robustcost}.  

\begin{theorem}[Convexification of Energy-Reserve Co-Optimization] \label{thm:Eq}
The energy-reserve co-optimization (\ref{eq:NONCVX}) becomes a linear program, if all SoC-dependent bids are EDCR bids and \eqref{eq:Simul_CD} can be relaxed.

In particular, the single-stage bid-in cost $f^*$ of the energy-reserve co-optimization of (\ref{eq:ROmileage}) is given by
\begin{subequations}
\begin{align}
\tilde{f}(q^c_t, q^d_t,e_t) := \underset{j\in [K]}{\rm max}\{\alpha_j(e_t) + c^d_{j} q^d_t-c^c_{j} q^c_t\}, \label{eq:CVX_single}\\
q^c_{t} := (p^c_{t}+\gamma^d_{t}r^d_{t})\tau,~~ q^d_{t} := (p^d_{t}+\gamma^u_{t}r^u_{t})\tau.\label{eq:gNotationn}
\end{align}
\end{subequations}
The multi-stage storage operation cost $F^*$ in (\ref{eq:robustcost}) has the following explicit convex piecewise linear  form
\beq
\begin{array}{lrl}\label{eq:ES_cost}
\tilde{F}(\qbf^c, \qbf^d;s):=\underset{j\in [K]}{\rm max}\{\alpha_j(s)+\sum_{t=1}^T(c^d_{j}q^d_t- c^c_{j}q^c_t)\}\\
\end{array}
\eeq
with $\alpha_j(s):=-\sum_{k=1}^{j-1}\frac{\Delta c^c_k(E_{k+1}-E_1)}{\eta}-\frac{c^c_{j}(s-E_1)}{\eta}+h(s)$ and $h(s):=\sum_{i=1}^K\mathbbm{1}_{\{s\in \Ec_i\}}(\frac{c^c_i(s-E_1)}{\eta}+\sum_{k=1}^{i-1}\frac{\Delta c^c_k (E_{k+1}-E_1)}{\eta})$. 
\end{theorem}

The proof is given in Appendix~\ref{sec:THM1proof}. 

As is stated in  Theorem~\ref{thm:Eq}, if all storage submit EDCR bids, we can transform the nontrivial robust optimization  (\ref{eq:NONCVX}) into a linear program. In the following, we give its convex piecewise linear program  equivalent formulation with the convexified storage cost \eqref{eq:ES_cost} to clear the SoC-dependent energy-reserve  market. The linear program reformulation is given in Appendix~\ref{sec:linearprogram}. 
   \vspace{-0.1in}
\begin{subequations}
\label{eq:CVX_bi}
\begin{align}
& \underset{\substack{\Omega}}{\rm minimize} && \sum_{i=1}^{M} (h_i(\gbf^e_i, \gbf_i^{u}, \gbf_i^{d})+ \tilde{F}_{i}(\qbf^c_i, \qbf^d_i;s_i)) \label{eq:objSysCost}\\
& \mbox{subject to}&& \forall t\in [T], \forall i\in [M],\nn \\
&~~~~\pmb{\mubf}[t]: && \pmb{S} (\pmb{g}^e[t]+\pbf^d[t]-{\pbf}^c[t]-\pmb{d}[t]) \le \pmb{l},\\
&~~~~\lambda_t: && {\bf 1}^\intercal(\pmb{g}^e[t]+\pbf^d[t]-{\pbf}^c[t]-\pmb{d}[t]) = 0,\\
&~~~~\beta^u_{t}: && \sum_i^M (g_{it}^{u} + r_{it}^u) \geq \xi_t^u, \\ 
&~~~~\beta^d_{t}: && \sum_i^M (g_{it}^{d} + r_{it}^d) \geq \xi_t^d, \\ 
&&& \eqref{eq:SoCevolveE}, \eqref{eq:SoClimit} ,\eqref{eq:gen}, \eqref{eq:SoCcons2}, \eqref{eq:SoCcons3}, \eqref{eq:gNotationn}, e_{i1}=s_i.\nn
    \end{align}
\end{subequations}

The following proposition supports the exact relaxation of bilinear constraint \eqref{eq:Simul_CD}.
\begin{proposition}[] \label{lemma:bidSpread}
If all SoC-dependent bids are EDCR bids and  LMPs from the energy-regulation co-optimization \eqref{eq:CVX_bi} are non-negative, the relaxation of the bilinear constraints $p^c_{it}p^d_{it}=0, \forall t\in [T], \forall i\in [M]$ in  (\ref{eq:CVX_bi}) is exact. 
\end{proposition}

See the proof in Appendix~\ref{sec:Prop1proof}. The non-negative assumption on LMP has been considered in \cite{Li18CSEEstorage, ChenBaldick21TPSstorageSCUCBinary} for the exact relaxation of bilinear constraint for differentiable objective functions.  Here we generalize it to a convex piecewise linear objective function via the subgradient measure  \cite[p. 281]{Rockafellar70convex}. The convex piecewise linear market clearing program in  \eqref{eq:CVX_bi} enables the operator to efficiently clear the electricity market with storage SoC-dependent costs. More precisely, the real-time market with rolling-window look-ahead dispatch \cite{Cong&Tong:22Allerton} can efficiently solve the market clearing problem \eqref{eq:CVX_bi} and keep rolling forward with updated forecasts for renewable and load.

\vspace{-0.15cm}
\subsection{ Pricing and profit in SoC-dependent market}

A significant benefit of convexification of the energy-reserve market is that it lends itself naturally to the locational marginal price (LMP) based pricing mechanism for energy and regulation capacity. Denote dual variables,  $\pmb{\mubf}[t]$, $\lambda_t$, $\beta^u_{t}$, and $\beta^d_{t}$ in \eqref{eq:CVX_bi}, and indicate the \emph{optimal} primal and dual solutions with superscript $*$.   The energy market clearing price is the LMP at  bus $i$ and interval $t$, given by
\beq\label{LMPDef}
\pi_{it}:=\lambda^*_t-\Sbf(:,i)^\intercal\mubf^*[t],
\eeq
where $\Sbf(:,i)$ is the $i$-th column of the shift factor matrix. The market clearing prices for the regulation up and regulation down capacities are given by  $\beta^{u*}_{t}$ and $\beta^{d*}_{t}$, respectively.

Based on the pricing of energy and regulation markets, storage  receives the market payment given by  
\beq\label{eq:payER}
{\cal P} = \sum_{t=1}^T(\pi_{t}(p^{d*}_{t} - p^{c*}_{t})+\beta^{u*}_{t}r^{u*}_{t}+\beta^{d*}_{t}r^{d*}_{t}).
\eeq

Based on SoC-dependent bid notations in Sec.~\ref{sec:SOCbid}, we parameterize the EDCR bid with $\thetabf:=(-\mathbf{c}^c, \mathbf{c}^d, \Ebf)$, and these parameters follow the EDCR condition in Definition~\ref{def:EDCR}. After storage submits EDCR  bid $\thetabf$ to the electricity market, the convex electricity market clearing problem (\ref{eq:CVX_bi}) produces the optimal energy-reserve dispatch $\phibf(\thetabf):=  (\pbf^{c*}, \pbf^{d*}, \rbf^{u*}, \rbf^{d*} )$, which is a function of storage  bidding parameters $\thetabf$. Similarly, the market payment to storage in \eqref{eq:payER} is a function of storage bidding parameters, \ie ${\cal P}(\phibf(\thetabf))$.

The bid-in cost of storage is denoted by  $\tilde{F}(\phibf(\thetabf); \thetabf, s)$   with formulation provided in  \eqref{eq:ES_cost}. Market payment  minus bid-in cost gives the {\em bid-in profit} of storage, defined by
\beq\label{EDCRProf}
\tilde{\Pi}(\thetabf) =  {\cal P}(\phibf(\thetabf)) - \tilde{F}(\phibf(\thetabf); \thetabf, s).
\eeq

Storage's bid-in cost $\tilde{F}$ from \eqref{eq:ES_cost} may have differences from its true cost denoted by $F$. Market payment minus the true cost of storage gives the {\em true profit} of storage,  defined by  
 \beq\label{eq:true cost}
\Pi(\thetabf) =  {\cal P}(\phibf(\thetabf))-F(\phibf(\thetabf); s).
\eeq

In the next section, we analyze storage bid-in profit and true profit under SoC-dependent  bids and SoC-independent bids.



%% file: OptEDCR_Bidnew_v6a.tex
Two questions are answered in this section: (i) How to generate the EDCR bid based on general storage cost data? (ii) Does the EDCR bid increase the profitability of BESS compared to the SoC-independent bid? The BESS profitability is influenced by many elements, including the method to generate storage bids, the market power of storage, and the reactions of other electricity market participants. Here,  we establish that, in a competitive electricity market, our EDCR bid generation approach can achieve a higher profitability than SoC-independent bids. Since we only consider one storage in this section, the storage index $i$ is dropped for simplicity.


\vspace{-0.15cm}
\subsection{ EDCR Bid Generation Approach}

Now that, $\thetabf:=(-\mathbf{c}^c, \mathbf{c}^d, \Ebf)$ is the notation for the EDCR bid, a given  SoC-independent bid is parameterized by  $\overline{\thetabf}:=(-\mathbf{1}\overline{c}^c, \mathbf{1}\overline{c}^d, \Ebf)$, representing the invariant bid-in price for charging and discharging power.\footnote{To avoid simultaneously charging and discharging behavior of storage under negative LMP, we assume $\overline{c}^c/\eta <\overline{c}^d$.} The following optimization generates an optimal EDCR bid $\thetabf^*$ with higher storage profitability under certain conditions. 
\begin{subequations}
\label{eq:OptEDCR_Bid}
\begin{align}
\thetabf^*(\overline{\thetabf}):= &\underset{\thetabf}{\rm arg~min} && \varphi(\thetabf)\\
& \mbox{subject to}&&  \eqref{eq:Ekk},\eqref{eq:a1}, \eqref{eq:EDCR},\nn\\
&&& \thetabf \le \bar{\thetabf},\label{cons:c1}\\
&&& \tilde{f}(\thetabf; \tilde{\sbf}) \geq \psi( \tilde{\sbf}), \forall \tilde{\sbf} \in {\cal S}.\label{cons:c2}
\end{align}
\end{subequations}
We define $\varphi(\thetabf):= ||\tilde{b}^c(\cdot|\thetabf)-b^c(\cdot)||_2^2+||\tilde{b}^d(\cdot|\thetabf)-b^d(\cdot)||_2^2$ in the objective. Storage SoC-dependent bids, $\tilde{b}^c(e_t)$ and $\tilde{b}^d(e_t)$, are defined in  (\ref{eq:SoCBid}).  Storage marginal cost and benefit are denoted by  $b^d(e_t)$ and $b^c(e_t)$ respectively, which can be computed from storage lab test data. This objective measures the approximation error  between the original storage marginal cost and the EDCR bid. With $N$  samples $(S_n, B^c_n, B^d_n)_{n=1}^N$ from the storage  marginal cost $b^d(e_t)$ and $b^c(e_t)$, we can compute  $\frac{1}{N}\sum_{n=1}^N((\tilde{b}^c(S_n|\thetabf)-B^c_n)^2+(\tilde{b}^d(S_n|\thetabf)-B^d_n)^2)$ for the objective function. 

Optimization \eqref{eq:OptEDCR_Bid} encodes the EDCR condition into its linear constraints  \eqref{eq:Ekk}, \eqref{eq:a1}, and \eqref{eq:EDCR}.\footnote{The strict inequality constraint in \eqref{eq:a1} is imposed by adding a small positive constant and reconstructing it into an inequality constraint. EDCR condition \eqref{eq:EDCR} is implemented by a linear constraint $c^c_{k+1}-c^c_{k}=\eta (c^d_{k+1}-c^d_{k})$ and optimization \eqref{eq:OptEDCR_Bid} has the SoC independent bid $\bar{\thetabf}$ as one feasible solution.}  Constraints \eqref{cons:c1}-\eqref{cons:c2} contain the core information for an EDCR to achieve higher profitability than a given SoC-independent bid. With \eqref{cons:c1}, we establish that the EDCR bid is bounded by the SoC-independent bid, \ie  
\beq \label{eq:bidparameter}
\bar{c}^c \le c^c_k\le c^d_k\le \bar{c}^d, \forall k \in [K],\eeq 

With \eqref{cons:c2}, the single-stage bid-in cost of storage is guaranteed to be no less than the true storage cost. Here, the initial SoC $s$ and the true SoC-dependent single-stage cost $\psi$ are given.  The analytical form of storage bid-in single stage cost $\tilde{f}(\thetabf; \tilde{\sbf})$ is given by \eqref{eq:CVX_single}, but here storage bid-in parameters $\thetabf$ are decision variables and storage actions $\tilde{\sbf}$ are given  parameters.\footnote{In  simulations, we relax  \eqref{cons:c2} and verify it for the optimal solution.} We denote $\Phi:=\{E_1, E_2,...,E_K\}$ as the set for all SoC breakpoints with $E_1=\underline{E}, E_K=\overline{E}$.  Here, we only consider limited storage single-stage charge and discharge actions. Storage single stage actions $\tilde{q}^c, \tilde{q}^d$ and SoC $\tilde{e}$ are collected in ${\cal S}:=\{ \tilde{\sbf}:= (\tilde{q}^c, \tilde{q}^d,\tilde{e}) | \tilde{e} \in \{s, \underline{E}, \overline{E}\}, \tilde{e} + \tilde{q}^c \eta -  \tilde{q}^d \in \Phi\} \subseteq   \mathbb{R}^3_+$. Related assumptions and analyses about the profitability of EDCR bids generated from \eqref{eq:OptEDCR_Bid} are explained in the following.
 
Note that optimization \eqref{eq:OptEDCR_Bid} is nonconvex. By fixing $\Ebf$ while solving for $(\mathbf{c}^c, \mathbf{c}^d)$, or fixing $(\mathbf{c}^c, \mathbf{c}^d)$ while solving for $\Ebf$, we can iteratively approach the (local) optimal solution by solving a convex problem in each iteration.

\vspace{-1.2em}
\subsection{ Profitability of  EDCR bid  }

We here assume storage  is a price-taker and has a small size. This means storage won't influence the market clearing prices by changing its bids, and the storage can only store limited energy. Such assumptions are typical in a competitive electricity market and can reduce the complexity when analyzing the profitability of EDCR bids. Note that such assumptions provide a sufficient but not necessary condition for storage to achieve a higher profit with the EDCR bid compared to the SoC-independent bid. In   Sec.~\ref{sec:Sim}, we show a more general empirical analysis without these restrictive assumptions. 

We've denoted $\phibf(\thetabf):= (\pbf^{c*}, \pbf^{d*}, \rbf^{u*}, \rbf^{d*})$ to include all storage optimal dispatch actions in the energy-reserve market. The following theorem analyzes the bid-in profit and true profit of storage under the EDCR bid and SoC-independent bid.
\begin{theorem}[Profitability of the EDCR Bid]\label{thm:ProfitEnergy}
 Let $\overline{\thetabf}$ be the SoC-independent bid and $\thetabf^*$ be the  EDCR bid  from  \eqref{eq:OptEDCR_Bid}. For a small-size price-taking storage, if $\phibf(\thetabf^*)  \neq  \phibf(\overline{\thetabf})$ and there's no dual degeneracy, then $\tilde{\Pi}(\thetabf^*) > \tilde{\Pi}(\overline{\thetabf})$ and  $\Pi(\thetabf^*) > \Pi(\overline{\thetabf})$.
\end{theorem}

Rigorous assumption statements for {\em small-size} and {\em price-taking} storage are shown in Appendix~\ref{sec:ProfAna} together with the  proof of Theorem~\ref{thm:ProfitEnergy}. When $\phibf(\thetabf^*) \neq \phibf(\overline{\thetabf})$, we have at least one time instance, such that storage actions under the EDCR bid are not the same as those under the SoC-independent bid. This essentially means that storage under the EDCR bid is scheduled more frequently than the SoC-independent bid, if storage is small-size, price-taking, and satisfies \eqref{cons:c1}. Thus, storage receives a higher market payment from the electricity market, resulting in a higher {\em bid-in profit}, \ie $\tilde{\Pi}(\thetabf^*) > \tilde{\Pi}(\overline{\thetabf})$ in a convex market clearing problem.

To achieve improvement for the {\em true profit} of storage, \ie $\Pi(\thetabf^*) > \Pi(\overline{\thetabf})$, \eqref{cons:c2} is the key. For small-size storage, ${\cal S}$ is a finite set including all potential  storage single-stage actions.  \eqref{cons:c2} guarantees that the single-stage bid-in cost of storage is always no less than the true storage cost.  That way, the storage true profit is always no less than the bid-in profit.

Note that the feasibility of  \eqref{eq:OptEDCR_Bid} is enough to guarantee the profitability of EDCR bid in Theorem~\ref{thm:ProfitEnergy}. But the optimality of \eqref{eq:OptEDCR_Bid} has the potential to further increase storage profit, which is empirically shown in the following sections and through extra simulations in Appendix \ref{sec:EMsim}.

%% file: Example_v4.tex
\begin{figure}[h]
\centering
\includegraphics[scale=0.3]{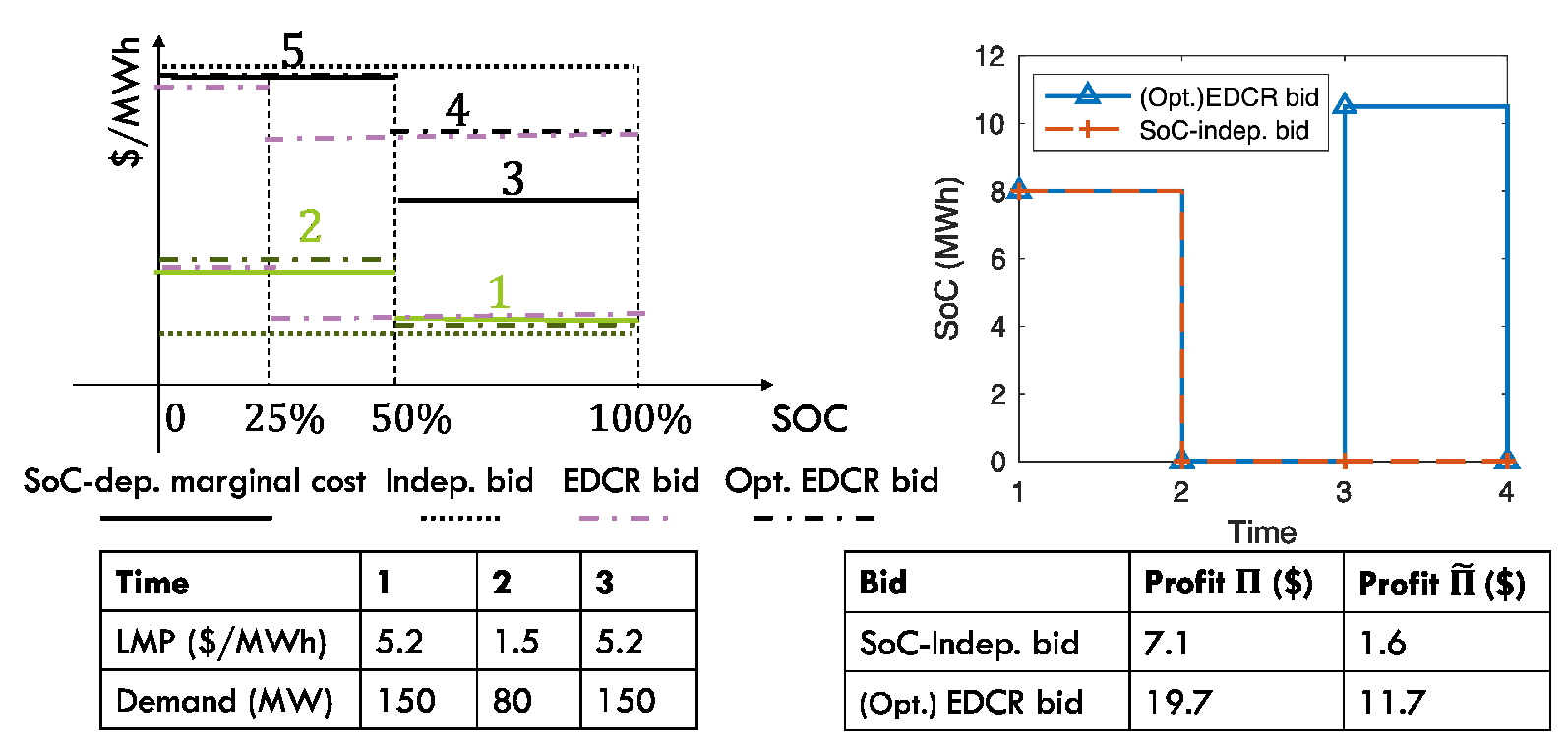} 
\caption{\scriptsize Storage bids and simulation results of the toy example, including market clearing prices, bid-in profitof storage, and true profit of storage. }
\label{fig:ToyEx}
\end{figure}

Consider a toy example with two generators, one inelastic demand, and an ideal storage\footnote{The round trip efficiency for ideal storage equals 1.} in the energy market with $T=3$. The bid-in marginal costs for the two generators are \$1.5/MWh and \$5.2/MWh, with capacities of 100MW and 1000MW, respectively. The storage SoC initializes at $s=8$ MWh, and the storage bids, which include the SoC-dependent marginal cost, EDCR bid, Opt.EDCR bid, and SoC independent bid, are shown in the top left of Fig~\ref{fig:ToyEx}. Here, SoC-dependent marginal costs  $b^d(e_t)$ and $b^c(e_t)$ are given by solid lines. We searched for  a feasibility solution of \eqref{eq:OptEDCR_Bid} when computing the EDCR bid. Meanwhile, the Opt.EDCR bid and SoC-independent bid came from the optimal solution of  \eqref{eq:OptEDCR_Bid}, minimizing the functional distance to the SoC-dependent marginal cost. 

For the  SoC-independent bid, we have $K=1$, $\overline{c}^d=5$ {\rm \$/MWh}, and $\overline{c}^c=1$ {\rm \$/MWh}.  For the  EDCR bid, we have $K=2$, $\cbf^d=(5, 4)$ {\rm \$/MWh}, $\cbf^c=(2, 1)$ {\rm \$/MWh}, $\Ebf=(0, 0.25, 1)\cdot \overline{E}$, and $\overline{E} = 10.5$ MWh. The  Opt. EDCR bid has  $\Ebf=(0, 0.5, 1)\cdot \overline{E}$. 
From the top right of Fig~\ref{fig:ToyEx}, it can be observed that  storage takes  the same actions under the SoC independent bid  and the EDCR bid at $t=1$. Meanwhile, storage with the EDCR bid has extra actions when $t=2, 3$, making it more profitable than the SoC independent bid. The bottom of Fig~\ref{fig:ToyEx} shows  LMP,  bid-in profit of storage $\tilde{\Pi}$, and true profit of storage $\Pi$. Compared with the SoC-independent bid, storage receives a higher bid-in profit and true profit with the EDCR bid.\footnote{The EDCR bid and Opt. EDCR bid receive the same simulation results in this section.} This validates Theorem~\ref{thm:ProfitEnergy}. 

%% file: Simulation_v6a.tex
We simulated energy-reserve co-optimization with one-shot dispatch for the day-ahead market-clearing and the rolling-window dispatch for the real-time market-clearing. More simulations for the pure energy market or reserve market can be found in Appendix \ref{sec:EMsim} and \cite{LiChenTong22SoC}, respectively.

\vspace{-1em}
\subsection{Parameter setting}
We considered 20 generators with energy bids ranging from \$0/MWh to \$246/MWh, regulation up bids from \$0/MWh to \$123/MWh, and regulation down bids from \$0/MWh to \$67/MWh. Generator capacity limits ranged from 150 MW to 1000 MW for energy dispatch, and from 10 MW to 200 MW for regulation capacity. Among the generators, there was one random solar generator with capacity limits of 500 MW  for energy dispatch and 10 MW for regulation capacity. We included one ideal energy storage with a capacity of 10.5 MWh.  We set energy demand and regulation requirements based on data from ISO New England\cite{ISONE:21} and PJM \cite{PJM:DataMiner}. We drew random scenarios from energy demands and solar output and computed the average metrics in the following simulation for the energy-reserve joint market. When computing the profit of BESS, payments from energy capacity and regulation capacity were considered. The regulation mileage payment belonging to ex-post settlements was separated from this simulation.\footnote{From the simulation result, storage with SoC-dependent bids had more regulation capacities cleared to follow the AGC signals and receive higher regulation mileage payments.} Details about units' bidding parameters and the procedure to generate random scenarios are provided in Appendix \ref{sec:simparamLoad}.  


We compared three categories of storage bidding parameters cleared in the energy-reserve market: (i) the  mixed integer program (MIP)  in \cite{ZhengXu22energy} was extended to simulate general SoC-dependent marginal cost, (ii) the SoC-dependent bid was cleared based on the current market clearing model in CAISO \cite{CAISO_StorageBid:21}, and (iii) the EDCR bids were cleared based on our proposed model in \eqref{eq:CVX_bi}. The parameters for storage, including SoC-dependent marginal cost, EDCR bid, Opt. EDCR bid, and SoC independent bid, are shown in the top left of Fig.~\ref{fig:ToyEx}. Details about the method to generate these bids are explained in Sec.~\ref{sec:Ex}. Here, we multiplied storage bidding parameters with a varying bid scale factor\footnote{The bid scale factor $\nu \in \mathbb{R}$ is shown in the x-axis of Fig.~\ref{fig:storage profit}-\ref{fig:RW}, which was implemented by $(\nu\mathbf{c}^c, \nu\mathbf{c}^d)$ for parameters $(\mathbf{c}^c, \mathbf{c}^d)$ in the top left of Fig~\ref{fig:ToyEx}.} to simulate different cases. 

 

\subsection{Day-ahead electricity market}
\label{Sec:oneshotsim}

We considered one-shot dispatch for the day-ahead electricity market with $T=7$. The initial SoC of the storage was 2.5 MWh. The regulation requirements were set one-sided at each time interval to avoid simultaneous clearing regulation up and down capacities. That way, the MIP proposed in \cite{LiChenTong22SoC} can accurately calculate the storage SoC-dependent  cost. We used this MIP to conduct the market clearing for the SoC-dependent marginal cost (yellow lines in Fig.~\ref{fig:ToyEx}) and calculated the true profit $\Pi$ for storage after obtaining the dispatch results under different storage bidding parameters. In Appendix \ref{sec:Prop2proof}, we explain the condition for clearing unidirectional regulation capacities. As for the EDCR bid and SoC-independent bid, we cleared with  convex market clearing problems \eqref{eq:CVX_bi}. 

\begin{figure}[!htb]
 \centering
 \vspace{-0.1in}
\includegraphics[scale=0.345]{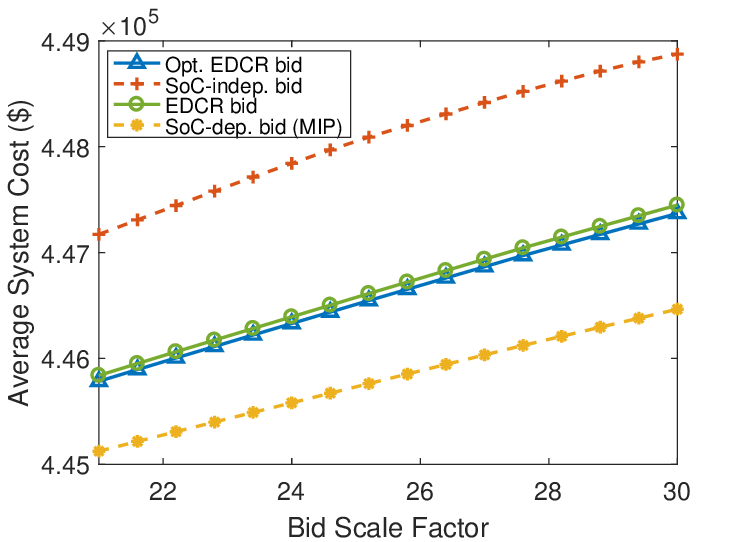}  
\includegraphics[scale=0.345]{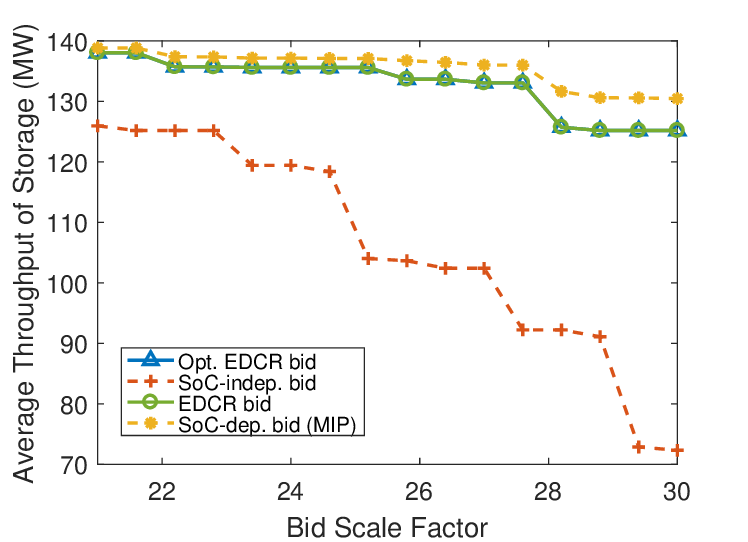}  
\includegraphics[scale=0.345]{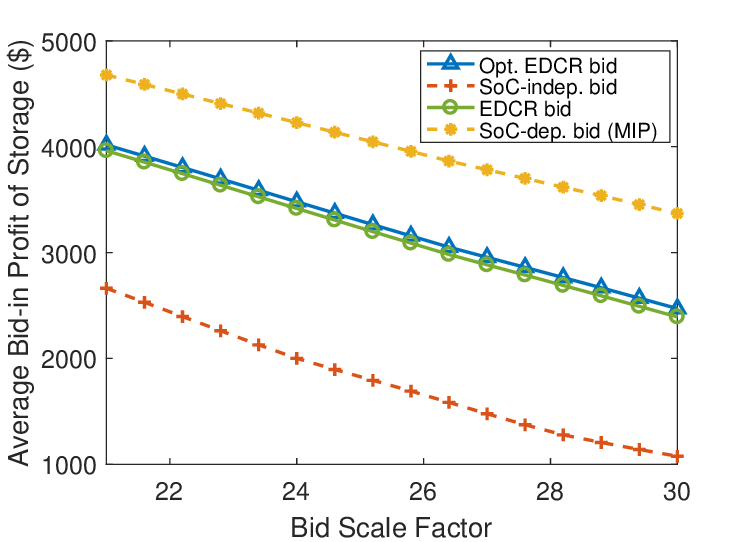}  
\includegraphics[scale=0.345]{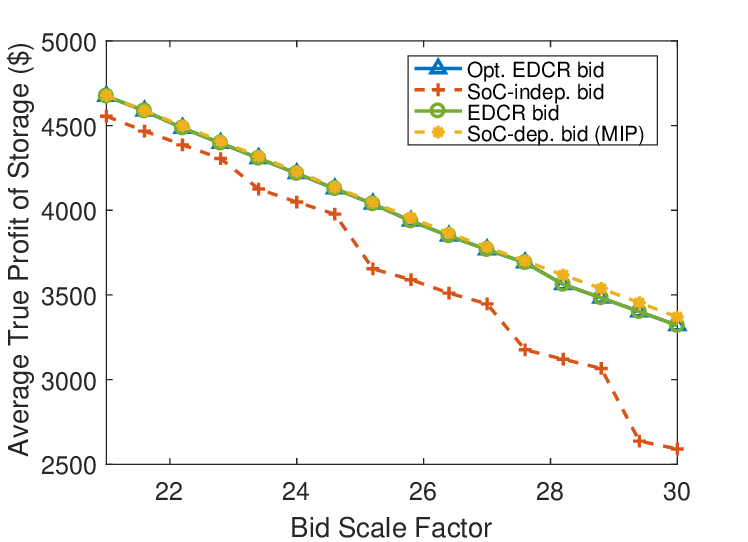}    
\caption{\scriptsize One-shot dispatch results. (Top left: system operation cost; top right: storage throughput; bottom left: storage bid-in profit; bottom right: storage true profit).}
\label{fig:storage profit}
\vspace{-0.1in}
\end{figure}

We simulated 500 random scenarios and calculated the average system cost computed by \eqref{eq:objSysCost}, average storage throughout,\footnote{Storage throughout is the sum of the energy dispatch and regulation capacities cleared over $T$-interval.} and average storage profits. These metrics versus the bid scale factor  were plotted in Fig.~\ref{fig:storage profit} for cases with different storage bidding parameters. The main conclusions are as follows.

First,  storage was used more frequently and received higher profits when submitting SoC-dependent bids rather than SoC-independent bids. As is shown by the top right of Fig.~\ref{fig:storage profit}, storage submitting the SoC-independent bid had the least throughput, representing a less frequent storage usage. Meanwhile, both bid-in profit $\tilde{\Pi}$ and true profit $\Pi$ of storage were higher when storage submitted the EDCR bid, which was consistent with Theorem~\ref{thm:ProfitEnergy}. Compared with the SoC-independent bid, the EDCR bid increased the bid-in profit of storage by 50\% to 130\%, and the true profit of storage by up to 28.1\%. 

Second, the EDCR bid achieved bidding effectiveness close to the original SoC-dependent marginal cost. The latter was computationally expensive and solved with MIP, bringing the issue of pricing nonconvex market and storage deficits (shown in the top right panel of Fig. 3 in \cite{LiChenTong22SoC}). In the right panels of Fig.~\ref{fig:storage profit}, compared with the original SoC-dependent storage costs in yellow, the EDCR bid had slightly lower true storage profits and utilization throughput due to the approximation error. This means enforcing the EDCR condition didn't cause much loss of storage profitability in this simulation. Despite the significant reduction in the computational burden imposed by MIP, EDCR bids effectively approximated the true cost of storage. We also noticed that the Opt.EDCR bid (blue curve) exhibited slightly better storage profit than the general EDCR bid (green curve). Extra simulations in Appendix \ref{sec:EMsim} show parallel phenomenons.

Third, system operation costs were reduced in EDCR bid cases because the storage's bids were closer to its true SoC-dependent marginal cost compared with the SoC-independent bid. For instance, when $\nu=21$, using the EDCR bid, the system cost decreased by approximately 0.3\% compared to the SoC-independent bid case.

Furthermore, when the bid scale factor increased, all metrics showed that the storage was less likely to be cleared in the energy-reserve co-optimization  because the storage bid was less competitive compared to the bid-in price of other generators.





\subsection{Real-time electricity market}

The convex market clearing brings more benefits to the rolling window dispatch in the real-time market, since it reduces the computation burden when considering SoC-dependent bids for storage. In \cite{Cong&Tong:22Allerton}, we give detailed explanations of the rolling-window dispatch and pricing. Here, we analyzed the market clearing results with different storage bidding parameters.

\begin{figure}[!htb]
   \centering
   \vspace{-0.1in}
\includegraphics[scale=0.34]{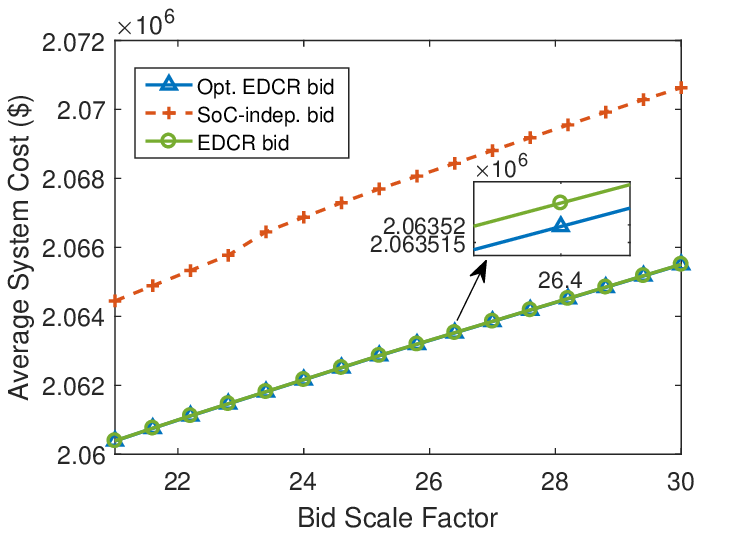}  
\includegraphics[scale=0.34]{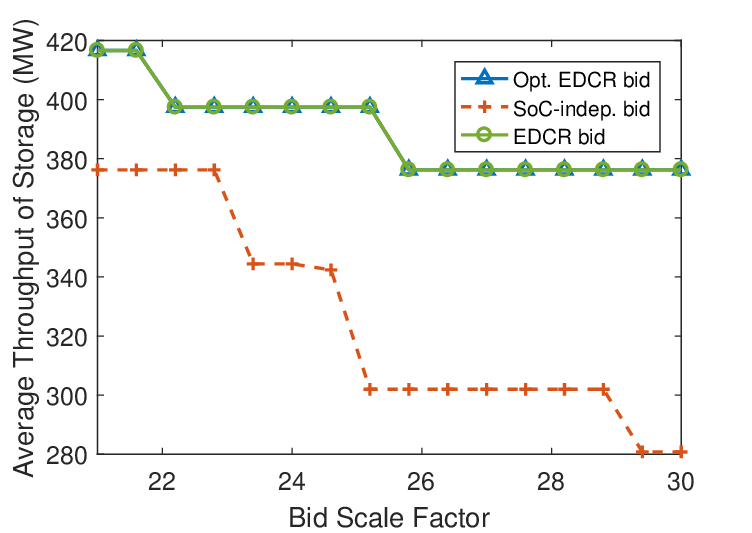}  
\includegraphics[scale=0.34]{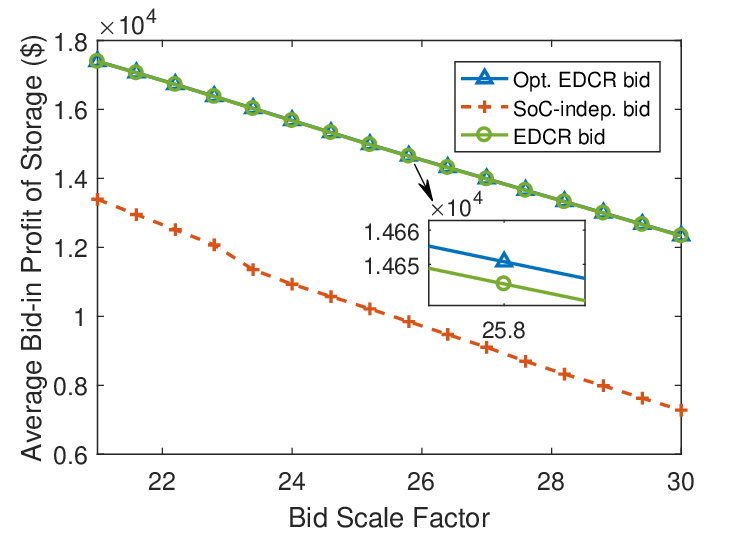}    
\caption{\scriptsize Rolling-window dispatch results. (Top left: system operation cost; top right: storage throughput; bottom: storage bid-in profit).}
\label{fig:RW}
\end{figure}

The initial storage SoC was 5 MWh. The simulation included $21$ rolling windows, and each rolling window had $T=4$ for the look-ahead horizon. As depicted in Fig.~\ref{fig:RW}, we compared the average system cost, storage throughput, and storage bid-in profit under 100 random scenarios. 

The main observations in Sec.~\ref{Sec:oneshotsim} still hold here.
The system cost was lower, and the storage throughput and bid-in profit were higher when storage submitted EDCR bids compared to the SoC-independent bid. In our parameter setting, the system cost decreased by up to 0.3\%, storage throughput was 10\%-30\% higher, and storage obtained approximately 1.5 times the bid-in profit when bidding EDCR bids rather than SoC-independent bids in the real-time energy-reserve co-optimized market.



%% file: ExCapacityReg_v2.tex
\subsection{Toy example of storage SoC-dependent cost}\label{sec:Ex_SOCdep}

We consider a toy example involving $T=1$ hour for the scheduling interval and $J=2$ for the AGC signal delivery intervals within  $T$. The capacity of the storage is set to be 10 MWh, and its bids are shown in the top left of Fig.~\ref{fig:ToyEx}. Assume that the storage initializes at 50\% SoC ($s=$5 MWh) and is cleared for both regulation up and down capacities of 1MW in $t=1$ hour.

Following a possible sequence of AGC signals, the storage first performs regulation up and then regulation down. The SoC trajectory is $[\hat{e}_{0,1},\hat{e}_{1,1},\hat{e}_{2,1}]=[5,4,5]$ MWh. In this case, the SoC-dependent storage cost for regulation up and down is \$5 and \$2, respectively, resulting in a total cost of \$7. Meanwhile, the SoC-independent cost is \$6.

Another possible sequence of AGC signals results in the storage regulating down first and then regulating up. The SoC trajectory is $[\hat{e}_{0,1},\hat{e}_{1,1},\hat{e}_{2,1}]=[5,6,5]$ MWh in this case. As a result, the SoC-dependent storage cost for regulation up and down is \$1 and \$3, respectively, a total cost of \$4. Meanwhile, the SoC-independent cost remains \$6 for this SoC trajectory.

Therefore, when the cleared regulation capacities are the same, the storage incurs the same bid-in cost for the SoC-independent bid but different costs for the SoC-dependent bid because the SoC trajectories are not identical.


%% file: THM1proof_v5a.tex
\subsection{Proof of Theorem~\ref{thm:Eq}}\label{sec:THM1proof}

We first introduce Lemma~\ref{lemma:Eq}, and then prove Theorem~\ref{thm:Eq}.

In Lemma~\ref{lemma:Eq}, we only analyze the storage cost for a single 15-min interval with index $t$. The initial SoC $e_t$ at the beginning of this 15-min interval is given. The SoC at the end of this 15-min interval is $e_{t+1}$.  We denote index $n$ and $m$ respectively below with $e_{t} \in {\cal E}_m$  and $e_{t+1} \in {\cal E}_n$. 
\begin{lemma}[]\label{lemma:Eq}
If storage's bid-in parameters satisfy the EDCR condition, then under Assumption~\ref{assume:single}, the  storage cost in (\ref{eq:SoCcostMile}) is piecewise linear convex given by
\begin{equation}
\begin{array}{lrl}\label{eq:EScostEDCR}
f(\mbf^u_t,\mbf^d_t, e_t)\\
=\underset{j\in [K]}{\rm max}\{\alpha_j(e_t) + c^d_{j}(p_t^dJ+\mathbf{1}^\intercal \mbf^u_t)\delta-c_j^c(p_t^cJ+\mathbf{1}^\intercal \mbf^d_t)\delta\}\\
= c^d_{n}(p_t^d\tau+\mathbf{1}^\intercal \mbf^u_t\delta)-c^c_{n}(p_t^c\tau+\mathbf{1}^\intercal \mbf^d_t\delta)\\
~~~+\begin{cases}
\sum_{k=m}^{n-1}\frac{-\Delta c^c_{k}}{\eta}(E_{k+1}-e_t), & n>m \\
0, & m=n  \\
\sum_{k=n+1}^{m} \Delta c^d_{k-1} (E_{k}-e_t), & n<m
\end{cases}
\end{array}
\eeq
with $\alpha_j(e_t):=-\sum_{k=1}^{j-1}\frac{\Delta c^c_k(E_{k+1}-E_1)}{\eta} - \frac{c^c_{j}(e_t-E_1)}{\eta}+h(e_t)$ and $h(e_t):=\sum_{i=1}^K\mathbbm{1}_{\{e_t\in \Ec_i\}}(\frac{c^c_i(e_t-E_1)}{\eta} + \sum_{k=1}^{i-1}\frac{\Delta c^c_k (E_{k+1}-E_1)}{\eta})$.
\end{lemma} 

Proof: The proof follows that of Theorem 1 in \cite{ChenTong22arXivSoC}, with the distinction that, in addition to the charging/discharging power in the energy market, we also calculate the operation cost caused by regulation up/down capacities in the regulation market, which is determined through regulation mileages. For the time index, from the definition of $\tau, \delta$ and $J$, we know that $\tau = J \cdot \delta$. 

Setting $\hat{e}_{\mbox{\tiny j+1}}=e_t+(\sum_{i=1}^{j}(p^c_t+m^d_{i,t})\eta-\sum_{i=1}^{j}(p^d_t+m^u_{i,t}))\tau$, the detailed proof corresponds to the methodology outlined in our previous paper \cite{ChenTong22arXivSoC}.   \QED

In our previous paper \cite{ChenTong22arXivSoC}, the EDCR condition for the energy market is proposed. Here, in parallel, we establish Lemma~\ref{lemma:Eq} for the joint energy and regulation markets.  \hfill 

Proof of Theorem~\ref{thm:Eq}:
In the upper-level model \eqref{eq:NONCVX}, the storage SoC is determined for each 15-min interval, with the values $e_t$ for $t\in [T]$. To ensure that the SoC $\hat{e}_{J+1,t}$ at the end of interval $t$ reaches the value $e_{t+1}$, the constraints defined in \eqref{eq:RegM} should be satisfied. Plugging $\mathbf{1}^\intercal \mbf^{u*}\delta =  \gamma^u_tr^u_t\tau$ and $\mathbf{1}^\intercal \mbf^{d*}\delta = \gamma^d_tr^d_t\tau$ into (\ref{eq:EScostEDCR}) results in the optimal objective value of \eqref{eq:ROmileage} as follows.
\begin{subequations}
\begin{align}
    & f^*(p^c_t, p^d_t, r^u_t, r^d_t,e_t)\nn \\
    &~= \underset{j\in [K]}{\rm max}\{\alpha_j(e_t) + c^d_{j}(p_t^d+\gamma^u_tr^u_t)\tau  -c^c_{j} (p^c_t +\gamma^d_tr^d_t)\tau\}\label{eq:objL} \\
    &~= c^d_{n}(p_t^d+\gamma^u_tr^u_t)\tau -c^c_{n} (p^c_t +\gamma^d_tr^d_t)\tau\nn\\
&~~+\begin{cases}
\sum_{k=m}^{n-1}\frac{-\Delta c^c_{k}}{\eta}(E_{k+1}-e_t), & n>m, \\
0, & m=n,  \\
\sum_{k=n+1}^{m} \Delta c^d_{k-1} (E_{k}-e_t), & n<m.
\end{cases}
\end{align}
\end{subequations}
Plug in the definition for $q^c_t, q^d_t$ from \eqref{eq:gNotationn} into \eqref{eq:objL}, we achieve the analytical optimal solution \eqref{eq:CVX_single} for  the worst-case single stage storage cost in (\ref{eq:ROmileage}).

We know that the initial SoC at $t=1$ is given, \ie $e_1=s$. Denote index $n$ and $m$ below respectively with initial SoC $e_{1} \in {\cal E}_m$  and the end-state SoC $e_{T+1} \in {\cal E}_n$. The multi-stage SoC-dependent cost  \eqref{eq:robustcost} can be computed by 
\beq
\begin{aligned}\label{eq:proofTHM1}
&F^*(\pbf^c, \pbf^d, \rbf^u, \rbf^d;s) :=\sum_{t=1}^T f^*(p^c_t, p^d_t, r^u_t, r^d_t,e_t)\\
&\overset{(a)}{=} \sum_{t=1}^T f(\mbf^{u*}_t,\mbf^{d*}_t)\\
&\overset{(b)}{=}\sum_{t=1}^T \underset{j\in [K]}{\rm max}\{\alpha_j(e_t) + c^d_{j}(p_t^dJ+\mathbf{1}^\intercal \mbf^{u*}_t)\delta-c_j^c(p_t^cJ+\mathbf{1}^\intercal \mbf^{d*}_t)\delta\}
\\
&\overset{(c)}{=}  \sum_{t=1}^T \underset{j\in [K]}{\rm max}\{\alpha_j(e_t) + c^d_{j}(p_t^d+\gamma^u_tr^u_t)\tau -c^c_{j} (p^c_t +\gamma^d_tr^d_t)\tau\}\\
&\overset{(d)}{=} \sum_{t=1}^T c^d_{n}(p_t^d+\gamma^u_tr^u_t)\tau-c^c_{n} (p^c_t +\gamma^d_tr^d_t)\tau\\
&~~~~~~~~+\begin{cases}
\sum_{k=m}^{n-1}\frac{-\Delta c^c_{k}}{\eta}(E_{k+1}-s), & n>m \\
0, & m=n  \\
\sum_{k=n+1}^{m} \Delta c^d_{k-1} (E_{k}-s), & n<m
\end{cases} \\
&\overset{(e)}{=}\underset{j\in [K]}{\rm max}\{\alpha_j(s)+\sum_{t=1}^T(c^d_{j}q^d_t- c^c_{j}q^c_t)\},
\end{aligned}
\eeq
where (a) comes from the definition in \eqref{eq:ROmileage}, (b) comes from \eqref{eq:EScostEDCR} in Lemma~\ref{lemma:Eq}, (c) comes from \eqref{eq:RegM}. (d) sums up \eqref{eq:objL} and can be proved by the same induction method in the proof of Theorem 1 in \cite{ChenTong22arXivSoC}. Note that, a difference between the proof of \cite{ChenTong22arXivSoC} and this paper is to use different SoC evolving constraint $
e_{t+1}=e_{t} +\eta q^c_t\tau-q^d_t\tau$ from \eqref{eq:SoCevolvePP}, which clears both the energy and regulation markets. (e) follows the proof of Theorem 1 in \cite{ChenTong22arXivSoC} showing that 
\beq
\begin{aligned}
    n=\underset{j\in [K]}{\rm arg~max}\{\alpha_j(s) + \sum_{t=1}^T(c^d_{j}q^d_t- c^c_{j}q^c_t)\}.
\end{aligned}
\eeq


Then we prove that \eqref{eq:ES_cost} is a convex function by showing it uses operations preserving convexity.
We know  $\alpha_j(s)$ is a constant, \eqref{eq:ES_cost} is adopting pointwise maximum operation over linear functions, so \eqref{eq:ES_cost} is  convex. 

Therefore, the objective function of the energy-reserve co-optimization (\ref{eq:NONCVX}) is  convex. If \eqref{eq:Simul_CD} can be relaxed, all constraints of (\ref{eq:NONCVX}) are linear. So, (\ref{eq:NONCVX}) is a convex optimization with equivalent form \eqref{eq:CVX_bi}. \QED

%% file: EDCRBidEX_v1.tex
\subsection{Example of  storage opportunity cost and EDCR bid}\label{sec:EX_EDCR}

In Sec.~\ref{sec:model}, we comment that the storage opportunity cost in arbitrage can be represented by the EDCR bid. Here, we follow \cite{ZhengXu22energy, Zhou&etal:24ACM} and provide the detailed derivation for single storage. We omit the storage index $i$ for simplicity.

With stochastic LMP  $\{\pi_t\}_{t=1}^T$ from the energy market, the profit maximization over multi-interval is given by
\begin{subequations}
\label{eq:SO_storage}
\begin{align}
& \underset{\pbf^d, \pbf^  \in\mathbb{R}^T_+}{\rm minimize} && \mathbb{E}[\sum_{t=1}^{T} \pi_t(p^c_t-p^d_t)] \\
& \mbox{subject to}&& \forall t\in [T],\nn \\
&&&e_{t+1}=e_{t} +\eta p^c_{t}-p^d_{t},~~\underline{E}\le e_{t+1} \le \overline{E}, \label{eq:cons1}\\ 
&&& p^d_{t}\le \bar{p}^d, ~~p^c_{t}\le \bar{p}^c,~~e_{i1}=s_i,  ~~\eqref{eq:Simul_CD}.\label{eq:consN}
    \end{align}
\end{subequations}
In the objective function, storage pays when charging and gets paid when discharging to capture the profit in “buy-low-sell-high". In the constraint, we omit the regulation capacities and time granularity parameters, so the complicated storage SoC constraints like \eqref{eq:SoCevolveE} and \eqref{eq:SoClimit} are simplified here. Follow the stochastic dynamic program and we backward recursively solve a single interval optimization: 
\begin{subequations}
\label{eq:SO_DP}
\begin{align}
V_t(e_t, \pi_t) =~& \underset{p^d_t, p^c_t \in\mathbb{R}_+}{\rm minimize} && \pi_t(p^c_t-p^d_t) + \mathbb{E}[V_{t+1}(e_{t+1}, \pi_{t+1})|\pi_t] \\
& \mbox{subject to}&& \eqref{eq:cons1},\eqref{eq:consN},
    \end{align}
\end{subequations}
where the first term in the objective is the immediate cost and the second term is the future opportunity cost for storage. Denote the marginal opportunity cost as $v_t(e_t, \pi_t):=\frac{\partial V_t(e_t, \pi_t)}{\partial e_t}$. The marginal influences of charging and discharging power for the storage opportunity cost are given by 
\begin{subequations}\label{eq:MCD}
\begin{align}
&b_t^c(e_t):=\frac{\partial \mathbb{E}[V_{t}(e_{t}, \pi_t)|\pi_{t-1}]}{\partial p^c_{t-1}} =  \eta \mathbb{E}[v_{t}(e_{t}, \pi_t)|\pi_{t-1}],\\
&b_t^d(e_t):=-\frac{\partial \mathbb{E}[V_{t}(e_{t}, \pi_t)|\pi_{t-1}]}{\partial p^d_{t-1}}  =  \mathbb{E}[v_{t}(e_{t}, \pi_t)|\pi_{t-1}],
\end{align}
\end{subequations}
which represent storage marginal benefit and marginal cost at different SoC levels, respectively.
%
The expected marginal opportunity cost function is monotonically decreasing over storage SoC $e_t$ and can be approximated by samples using equation (4) of \cite{ZhengXu22energy}, which is a step function \ie $\mathbb{E}[v_{t}(e_{t}, \pi_t)|\pi_{t-1}] \approx \sum_{k=1}^K \tilde{v}_k\mathbbm{1}_{\{e_t\in \Ec_k\}}$. Follow notations from equation \eqref{eq:SoCBid} of this paper, SoC-dependent bids for charging and discharging power are step functions with 
\beq\label{eq:bidcheck}
\begin{aligned}
c^c_k = \eta \tilde{v}_k,~~ c^d_k =  \tilde{v}_k, \forall k \in [K].
\end{aligned}
\eeq
Here we can check that $\frac{c^c_{k+1}-c^c_{k}}{c^d_{k+1}-c^d_{k}}=\eta, \forall k \in [K-1].$ So, the EDCR condition in \eqref{eq:EDCR} is satisfied.

The SoC-dependent bid derived from \cite{ZhengXu22energy} won't satisfy the EDCR condition when the storage round trip efficiency and degradation cost parameters are SoC-dependent. In these cases, we can use the method in Sec.~\ref{sec:OptApprox} to compute the EDCR bid with the smallest distance to the original SoC-dependent bid.

%% file: LPReform_v0.tex
\vspace{-1.3em}
\subsection{linear program equivalent formulation of  \eqref{eq:CVX_bi}} \label{sec:linearprogram}
We can rewrite the convex piecewise linear SoC-dependent bid-in cost \eqref{eq:ES_cost} with a set of linear constraints given by
\beq
\upsilon \geq \alpha_j(s)+\sum_{t=1}^T(c^d_{j}q^d_t- c^c_{j}q^c_t), \forall j\in [K].\eeq

Add storage index $i$, and assume generator cost $h_i(\gbf^e_i, \gbf_i^{u}, \gbf_i^{d})$ is linear for simplicity. We denote $\upsilonbf[j]:= (\upsilon_{1, j},..., \upsilon_{i, j}, ..., \upsilon_{M, j})$, $\alphabf_j(\sbf) := (\alpha_{1, j}(s_1),..., \alpha_{i, j}(s_i), ..., \alpha_{M, j}(s_M))$, $\cbf^d[j]:= (c^d_{1, j},..., c^d_{i, j}, ..., c^d_{M, j})$ and $\qbf^d[t]:= (q^d_{1, t},..., q^d_{i, t}, ..., q^d_{M, t})$. Similar notations are defined for $\qbf^c[t]$ and $\cbf^c[j]$. Denote the component-wise multiplication (Hadamard product) with $\odot$. The linear program reformulation of  \eqref{eq:CVX_bi} is given by
\begin{subequations}\label{eq:CVX_biLPform}
\begin{align}
& \underset{\substack{\Omega\cup \{\upsilonbf[j]\}}}{\rm minimize} && \sum_{i=1}^{M} (h_i(\gbf^e_i, \gbf_i^{u}, \gbf_i^{d})+ \upsilon_{i})\\
& \mbox{subject to}&& \forall t\in [T], \forall i\in [M],\forall j\in [K],\nn \\
&&& \upsilonbf[j] \geq \alphabf_j(\sbf)+\sum_{t=1}^T(\cbf^d[j]\odot \qbf^d[t] - \cbf^c[j]\odot\qbf^c[t]),\\ 
&~~~~\pmb{\mubf}[t]: && \pmb{S} (\pmb{g}^e[t]+\pbf^d[t]-{\pbf}^c[t]-\pmb{d}[t]) \le \pmb{l},\\
&~~~~\lambda_t: && {\bf 1}^\intercal(\pmb{g}^e[t]+\pbf^d[t]-{\pbf}^c[t]-\pmb{d}[t]) = 0,\\
&~~~~\beta^u_{t}: && \sum_i^M (g_{it}^{u} + r_{it}^u) \geq \xi_t^u, \\ 
&~~~~\beta^d_{t}: && \sum_i^M (g_{it}^{d} + r_{it}^d) \geq \xi_t^d, \\ 
&&& \eqref{eq:SoCevolveE}, \eqref{eq:SoClimit} ,\eqref{eq:gen}, \eqref{eq:SoCcons2}, \eqref{eq:SoCcons3}, \eqref{eq:gNotationn}, e_{i1}=s_i.\nn
    \end{align}
\end{subequations}

%% file: PropProof_v3a.tex
\vspace{-1.3em}
\subsection{Proof of Proposition~\ref{lemma:bidSpread}}\label{sec:Prop1proof}
We complete the dual variables for \eqref{eq:CVX_bi} below and then write the proof.

\begin{subequations}
\label{eq:CVX_bicomplete}
\begin{align}
& \underset{\substack{\Omega}}{\rm minimize} && \sum_{i=1}^{M} (h_i(\gbf^e_i, \gbf_i^{u}, \gbf_i^{d})+ \tilde{F}_{i}(\qbf^c_i, \qbf^d_i;s_i)) \\
& \mbox{subject to}&& \forall t\in \{1,...,T\}, \forall i\in \{1,...,M\},\nn \\
&~~~~\pmb{\mubf}[t]: && \pmb{S} (\pmb{g}^e[t]+\pbf^d[t]-{\pbf}^c[t]-\pmb{d}[t]) \le \pmb{l},\\
&~~~~\lambda_t: && {\bf 1}^\intercal(\pmb{g}^e[t]+\pbf^d[t]-{\pbf}^c[t]-\pmb{d}[t]) = 0,\\
&~~~~\beta^u_{t}: && \sum_i^M (g_{it}^{u} + r_{it}^u) \geq \xi_t^u, \\ 
&~~~~\beta^d_{t}: && \sum_i^M (g_{it}^{d} + r_{it}^d) \geq \xi_t^d, \\
&~~~~\varrho_{it}: && e_{it} +\eta_i q^c_{it} -q^d_{it}   = e_{i(t+1)},\label{eq:socEEE}\\
&~~~~\omega^c_{it}: &&e_{it} +\eta_i q^c_{it}    \leq \bar{E_i},\\
&~~~~\omega^d_{it}: &&\underline{E_i} \le e_{it}-q^d_{it},\\
&&& g^e_{it}+g^u_{it}\leq \bar{g}_i,~~ \underline{g}_i \le g^e_{it}-g^d_{it}, \\ 
&(\underline{\vartheta}^d_{it},\bar{\vartheta}^d_{it}):&& 0\le r^d_{it}\le \bar{r}^d_i,\\
&(\underline{\vartheta}^u_{it},\bar{\vartheta}^u_{it}): && 0\le r^u_{it}\le \bar{r}^u_i,\\ 
&(\underline{\rho}^d_{it},\bar{\rho}^d_{it}): && 0 \le p^d_{it}\le \bar{p}^d_i,\\
&(\underline{\rho}^c_{it},\bar{\rho}^c_{it}): &&0 \le p^c_{it}\le \bar{p}^c_i,\\
&&&  q^c_{it} = (p^c_{it}+\gamma^d_{it}r^d_{it})\tau,\\
&&& q^d_{it} = (p^d_{it}+\gamma^u_{it}r^u_{it})\tau.
    \end{align}
\end{subequations}
Note that in constraint \eqref{eq:socEEE} with $t=1$, we directly replace in $e_{i1}=s_i$.

Proof: We prove Proposition~\ref{lemma:bidSpread} by contradiction. Assume that there exists an optimal solution with the simultaneous charging and discharging power, \ie $p^{c*}_{it}>0, p^{d*}_{it}>0$. After relaxing the constraint, $p^c_{it}p^d_{it}=0, \forall i, t$, (\ref{eq:CVX_bi}) is convex with a subdifferentiable objective when the  EDCR condition is satisfied. With dual variables defined in \eqref{eq:CVX_bicomplete} and the KKT conditions\footnote{ Write the Lagrangian function of \eqref{eq:CVX_bicomplete}, take derivative  over $p^c_{it}$  and $p^d_{it}$ respectively, do the operation to cancel the dual variable associate with (28n) (28o), and then we achieve \eqref{eq:EDCRKKTEE}.} \cite[p. 281]{Rockafellar70convex}, there exist $ \kappa^c_i \in \frac{\partial}{\partial p^c_{it}}  \tilde{F}_{i}(\qbf^c_i, \qbf^d_i;s_i) $ and $ \kappa^{d}_i \in \frac{\partial}{\partial p^{d}_{it}}  \tilde{F}_{i}(\qbf^c_i, \qbf^d_i;s_i) $, satisfying
\beq\label{eq:EDCRKKTEE}
\begin{cases} 
 \kappa^c_i +\pi_{it}+(-\varrho^*_{it}+\omega^{c*}_{it})\eta_i\tau-\underline{\rho}^{c*}_{it}+\bar{\rho}^{c*}_{it}=0,& \\
 \kappa^d_i-\pi_{it}+(\varrho_{it}^*+\omega^{d*}_{it})\tau-\underline{\rho}^{d*}_{it}+\bar{\rho}^{d*}_{it}=0, & 
\end{cases}
\eeq
\beq
\label{eq:combKKT}
\Rightarrow  \frac{\kappa^c _i}{\eta_i}+\kappa^d_i+\pi_{it}(\frac{1}{\eta_i}-1)+(\omega^{c*}_{it}+\omega^{d*}_{it})\tau+\frac{\bar{\rho}^{c*}_{it}}{\eta_i}+\bar{\rho}^{d*}_{it}=0,
\eeq
where $\pi_{it}$ is the LMP defined in  \eqref{LMPDef}, $\frac{1}{\eta_i}\geq 1, \omega^{c*}_{it}\geq 0, \omega^{d*}_{it}\geq 0, \bar{\rho}^{c*}_{it}\geq 0, \bar{\rho}^{d*}_{it}\geq 0$, and  we have $\underline{\rho}^{c*}_{it}=0, \underline{\rho}^{d*}_{it}=0$ from the complementary slackness conditions. 
The subgradient of  the storage cost function, $\frac{\partial}{\partial p^c_{it}}  \tilde{F}_{i}(\qbf^c_i, \qbf^d_i;s_i) $ and $\frac{\partial}{\partial p^{d}_{it}}  \tilde{F}_{i}(\qbf^c_i, \qbf^d_i;s_i)$, can be respectively computed by
\beq\label{eq:subgradientDef1}
\frac{\partial}{\partial p^{c}_{it}}  \tilde{F}_{i}(\qbf^c_i, \qbf^d_i;s_i) =\begin{cases} \{-c^c_{ik}\}\tau,& \mbox{if}~ q^{c}_{it}\in \mbox{Int}~\Ec_{ik},\footnote{The subgradient equals to $-c^c_{i1}\tau$ if $q^c_{it}=E_{i1}$, and equals to $-c^c_{i\mbox {\tiny K}}\tau$ if $q^c_{it}=E_{i \mbox {\tiny (K+1)}}$.}\\ 
[-c^c_{ik}\tau, -c^c_{i(k+1)}\tau], & \mbox{if}~ q^c_{it}=E_{i(k+1)}, \end{cases}\eeq

\beq\label{eq:subgradientDef2}
\frac{\partial}{\partial p^{d}_{it}}  \tilde{F}_{i}(\qbf^c_i, \qbf^d_i;s_i)=\begin{cases} \{c^d_{ik}\}\tau, & \mbox{if}~ q^d_{it}\in \mbox{Int}~\Ec_{ik},\\ 
[c^d_{i(k+1)}\tau, c^d_{ik}\tau], & \mbox{if}~ q^d_{it}=E_{i(k+1)}. \end{cases}\footnote{The subgradient equals to $c^d_{i1}\tau$ if $q^d_{it}=E_{i1}$, and $c^d_{i\mbox {\tiny K}}\tau$ if $q^d_{it}=E_{i \mbox {\tiny (K+1)}}$.}
\eeq 

So, under Definition~\ref{assume:single}, we have $ \frac{1}{\eta_i} \kappa^c_i +\kappa^d_i> 0$, which contradicts to the assumption that the LMP $\pi_{it}$ is nonnegative in equation (\ref{eq:combKKT}). \QED

%% file: Prop2Proof_v0.tex
\subsection{ Proposition~\ref{Prop:Simultaneous} and its proof}\label{sec:Prop2proof}

In \cite{LiChenTong22SoC}, we propose a heuristic approach to approximate the worst-case storage cost with a mixed integer program (MIP), which has no approximation error when unidirectional regulation capacities are cleared. Below we explain a sufficient condition for clearing  unidirectional regulation capacities.
 \begin{proposition}\label{Prop:Simultaneous} If all SoC-dependent bids are EDCR bids with bid-in parameters $(c_{ik}^c,c_{ik}^d)$ and $\eta_i$ satisfying
\beq\label{eq:ReqCondi}
(c^d_{iK}-c^c_{i1}/\eta_i)\tau>\beta^{u*}_t/\gamma^u_{it}+\beta^{d*}_t/(\eta\gamma^d_{it}),
\eeq 
then the storage $i$ has one-side regulation capacity cleared over all time intervals, \ie $r^{u*}_{it} r^{d*}_{it} = 0, \forall t\in [T], \forall i\in [M]$. 
\end{proposition}
 
Proof: Assume for the sake of contradiction that there exists an optimal solution in which storage $i$ simultaneously offers both regulation up and regulation down capacities at time interval $t$, \ie $r^u_{it}>0, r^d_{it}>0$. After the EDCR condition is satisfied, the market clearing model (\ref{eq:CVX_bi}) exhibits convexity with respect to the subdifferentiable objective function \eqref{eq:ES_cost}. With dual variables defined in \eqref{eq:CVX_bicomplete} and the KKT conditions, there exist subgradients $\iota^u_i\in \frac{\partial}{\partial r^u_{it}}  \tilde{F}_{i}(\qbf^c_i, \qbf^d_i;s_i)$ and $\iota^d_i \in  \frac{\partial}{\partial r^d_{it}}  \tilde{F}_{i}(\qbf^c_i, \qbf^d_i;s_i)$ such that
\beq\label{eq:EDCRKKTReg}
\begin{cases} 
 \iota^u_i+(\varrho^*_{it}+\omega^{d*}_{it})\gamma^u_{it}\tau-\beta^{u*}_t-\underline{\vartheta}_{it}^{u*}+\bar{\vartheta}_{it}^{u*}=0, & \\
 \iota^d_i+(-\varrho^*_{it}+\omega^{c*}_{it})\eta_i\gamma^d_{it}\tau-\beta^{d*}_t-\underline{\vartheta}_{it}^{d*}+\bar{\vartheta}_{it}^{d*} =0,& 
\end{cases}
\eeq
\beq
\begin{aligned}\label{eq:contradict}
\Rightarrow \frac{\iota^u_{i}-\beta^{u*}_t+\bar{\vartheta}_{it}^{u*}}{\gamma^u_{it}\tau}+\frac{\iota^d_{i}-\beta^{d*}_t+\bar{\vartheta}_{it}^{d*}}{\eta_i\gamma^d_{it}\tau}+\omega^{d*}_{it}+\omega^{c*}_{it}=0,
\end{aligned}
\eeq
where $\bar{\vartheta}_{it}^{u*}\geq 0, \bar{\vartheta}_{it}^{d*}\geq 0, \omega^{c*}_{it}\geq 0, \omega^{d*}_{it}\geq 0$, and  we have $\underline{\vartheta}_{it}^{u*}=0, \underline{\vartheta}_{it}^{d*}=0$ from the complementary slackness conditions.  From \eqref{eq:subgradientDef1}-\eqref{eq:subgradientDef2}, the subgradients $\iota^u_i$ and $\iota^d_i$ are equal to $c^d_{ik_1}\gamma^u_{it}\tau$ and $-c^c_{ik_2}\gamma^d_{it}\tau$ respectively for some $k_1,k_2\in [K]$. 
In Proposition~\ref{Prop:Simultaneous}, restrictions \eqref{eq:ReqCondi} are only imposed on $c_K^d$ and $c_1^c$. Combined \eqref{eq:ReqCondi} with the monotonicity of the SoC-dependent bid under Assumption~\ref{assume:single}, we have,  $\forall k_1,k_2\in  [K]$,
\beq
\begin{aligned}
(c^d_{iK}-c^c_{i1}/\eta_i)\tau>\beta^{u*}_t/\gamma^u_{it}+\beta^{d*}_t/(\eta_i\gamma^d_{it})\\
\Rightarrow (c^d_{ik_1}-c^c_{ik_2}/\eta_i)\tau>\beta^{u*}_t/\gamma^u_{it}+\beta^{d*}_t/(\eta_i\gamma^d_{it}).\nn
\end{aligned}
\eeq

Therefore, we have 
$\frac{\iota^u_i}{\gamma^u_{it}}+\frac{\iota^d_i}{\eta_i\gamma^d_{it}}>\frac{\beta^{u*}_t}{\gamma^u_{it}}+\frac{\beta^{d*}_t }{\eta_i\gamma^d_{it}} \Rightarrow $
\beq
\frac{\iota^u_{i}-\beta^{u*}_t+\bar{\vartheta}_{it}^{u*}}{\gamma^u_{it}\tau}+\frac{\iota^d_{i}-\beta^{d*}_t+\bar{\vartheta}_{it}^{d*}}{\eta_i\gamma^d_{it}\tau}+\omega^{d*}_{it}+\omega^{c*}_{it}>0,
\eeq
leading to a contradiction for \eqref{eq:contradict}. \QED

%% file: FinalProfitProof_v5a.tex
\subsection{Proof of Theorem~\ref{thm:ProfitEnergy} }\label{sec:ProfAna}
Since we only consider one storage in this section, the storage index $i$ is dropped for simplicity.

For a given SoC-independent bid $\overline{\thetabf}$, we can compute an EDCR bid $\thetabf^*$ with \eqref{eq:OptEDCR_Bid}. We make the following two assumptions when storage bids with $\thetabf^*$ and $\overline{\thetabf}$.  A1-A2 apply to small-size storage, which won't influence the market clearing prices by changing its bid-in parameters.
 
\begin{itemize}
     \item[A1:] Let $\chi$ be the time interval   that the storage takes the first charging/discharging action during a $T$-interval dispatch.  The optimal dispatch of storage  SoC has 
     \beq
     e^*_t \in \{\underline{E}, \overline{E}\}, \forall t \in \{ \chi+1, ..., T\}, ~\text{and}~ e^*_{T+1} \in  \Phi.\nn
     \eeq
      \item[A2:] When storage changes bidding parameters, electricity market clearing prices computed from \eqref{eq:CVX_bi} stay the same.
\end{itemize}    


Outline for the proof of Theorem~\ref{thm:ProfitEnergy}: Consider sequential charging and discharging optimal actions for storage with SoC independent bid happening at time ${\cal A}= \{t_1, t_2,...,t_i, ..., t_N\}$\footnote{Here, $t_1$ is the first action interval represented by $\chi$ in A1. $t_i$ is the time interval index for the $i$-th action. We consider $N$ actions for storage.} in the $T$-interval optimization horizon and ${\cal A } \subseteq [T]$. We here prove backward from $t_N$ to $t_1$ showing that when switching from SoC independent bid $\overline{\thetabf}$ to EDCR bid  $\thetabf^*$ satisfying constraints in \eqref{eq:OptEDCR_Bid}, the storage has the same optimal actions for time intervals in ${\cal A}$, and the storage may have extra actions in other time intervals $[T] \setminus {\cal A}$. Then, we show the true storage profit $\Pi$ and bid-in storage profit $\tilde{\Pi}$ in $[T] \setminus {\cal A}$ is nonnegative under the EDCR bid. Therefore, the true storage profit $\Pi$ and bid-in storage profit $\tilde{\Pi}$  over all time intervals $[T]$ under the EDCR bid $\thetabf^*$  is no less than that under the SoC-independent bid $\overline{\theta}$. 

Proof of Theorem~\ref{thm:ProfitEnergy}: Storage unit index $i$ is dropped in this section. When A1 is satisfied, the upper bounds  in constraints  \eqref{eq:SoCcons2} \eqref{eq:SoCcons3} and dual variables associated, \ie $\bar{\rho}^{c*}_{t}, \bar{\rho}^{d*}_{t}, \bar{\vartheta}^{d*}_{t}, \bar{\vartheta}^{u*}_{t}$,  can be ignored. For simplicity, we write the proof below in terms of energy market actions, $p^{c*}_t, p^{d*}_t$, and  LMP.\footnote{This means $\qbf^c = \pbf^c, \qbf^d = \pbf^d, \rbf^d = \rbf^u = \mathbf{0}.$} The same proof works for regulation market actions, $r^{d*}_t, r^{u*}_t$, and price, by replacing charging electricity price with $-\beta^{u*}_t$, discharging electricity price with $\beta^{d*}_t$, and KKT condition \eqref{eq:EDCRKKTEE} used below with \eqref{eq:EDCRKKTReg}.

\underline{At the time $t_N$}, A1 indicates that, under the SoC-independent bid $\overline{\theta}$, the storage at time $t_N$ can \underline{1) charge to $\overline{E}$, or 2) discharge to $\underline{E}$}. When the storage switches to EDCR bid, we here prove the storage action stays the same at time $t_N$. 

In \underline{case 1) charge to  $\overline{E}$},  suppose the storage receives LMP $\pi_{t_N}$ and it submits SoC-independent bid $(\bar{c}^c, \bar{c}^d)$. When the objective function is the SoC-independent bid, the KKT condition of (\ref{eq:CVX_bicomplete}) gives 
\beq\label{eq:KKTeqcharge}
 -\bar{c}^c +\pi_{t_N} +(-\varrho^*_{t_N}+\omega^{c*}_{t_N})\eta \tau-\underline{\rho}^{c*}_{t_N}=0.
\eeq
LMP is defined in \eqref{LMPDef}. At time $t_N$, from the complementary slackness condition, we have $\underline{\rho}^{c*}_{t_N}=0$, $\omega^{c*}_{i+1}\geq 0$, $\omega^{d*}_{i+1} =0$, $\forall i \geq t_N$. With    Lemma~\ref{eq:SoCLemma}, we have
\beq\label{eq:KKTeqC}
 -\bar{c}^c +\pi_{t_N} -\sum_{n=t_N}^T(0-\omega_n^{c*})\eta \tau = 0, 
\eeq
which indicates $\pi_{t_N} \le \bar{c}^c$. 

When the storage switches to EDCR bid, we know LMP stays the same from A2. So we can get KKT condition under EDCR bid like \eqref{eq:EDCRKKTEE} via increasing $\omega^{c*}_{t_N}$ in \eqref{eq:KKTeqC} by $\Delta \omega^{c*}_{t_N} =  (-\kappa^c - \bar{c}^c)/(\eta\tau) $, and replacing SoC independent bid parameter $\bar{c}^c$ with the subgradiant $-\kappa^c$ defined in \eqref{eq:subgradientDef1}. From the constraints in \eqref{eq:OptEDCR_Bid}, we know $- \bar{c}^c \geq \kappa^c$, thus $\Delta \omega^{c*}_{t_N}  \geq 0$. After $t_N$, if the storage with EDCR bid has another discharge action at interval $k$ under EDCR bid, \ie $\omega^{d*}_k\geq 0, k>t_N+1$, then we can increase $\omega^{c*}_{t_N}$ to cover this part, \ie $\Delta \omega^{c*}_{t_N} =  (-\kappa^c - \bar{c}^c)/(\eta\tau) + \omega^{d*}_k \geq 0$. So the optimal storage dispatch under EDCR bid $\thetabf^*$ charges to $\overline{E}$ at time $t_N$.  

For \underline{case  2) discharge to $\underline{E}$}, similar proof can be applied.

\underline{Among time interval $[t_{N-1}, t_N]$}, A1 indicates that, under the SoC-independent bid, the storage can \underline{1) charge to $\overline{E}$ at time $t_{N-1}$ and discharge to $\underline{E}$  at time $t_{N}$} or \underline{2) discharge to $\underline{E}$ at time $t_{N-1}$ and charge to $\overline{E}$ at time $t_{N}$}.

In case \underline{1) charge to $\overline{E}$ at time $t_{N-1}$ and discharge to} \underline{$\underline{E}$  at time $t_{N}$} , we denote $i=t_{N-1}$ and $j=t_{N}$ for the following equations. From Lemma.~\ref{lemma:CoupleCD}, we have
\beq
\begin{array}{l}
\pi_{i} -\bar{c}^c-\eta (\pi_{j} -\bar{c}^d) \le 0
\end{array}
\eeq 
From constraints in \eqref{eq:OptEDCR_Bid},
\beq \pi_{i} +\kappa^c-\eta (\pi_{j} -\kappa^d) \le \pi_{i} -\bar{c}^c-\eta (\pi_{j} -\bar{c}^d)
\eeq 
So if the storage submits EDCR bid and it doesn't has extra dispatches among time interval $(t_{N-1}, t_N)$, we can get KKT condition under EDCR bid like \eqref{eq:EDCRKKTEE} by replacing SoC independent bid parameter $\bar{c}^c$ and $\bar{c}^d$ with the subgradiant $-\kappa^c$ and $\kappa^d$ respectively, and increasing $ \omega^{c*}_{i}$ in \eqref{eq:KKTeq2} to a certain value. Intuitively, we can increase $ \omega^{c*}_{i}$ by  $\Delta  \omega^{c*}_{i} =  (-\kappa^c-\bar{c}^c)/(\eta\tau)$. However, since the value of $ \omega^{d*}_{j}$ has been adjusted, the increment for $ \omega^{c*}_{i}$ should be $ \Delta  \omega^{c*}_{i} = (-\kappa^c-\bar{c}^c)/(\eta\tau)+(\bar{c}^d-\kappa^d)/\tau\geq 0$. The optimal storage dispatch under EDCR bid also charges $\overline{E}$ at time  $i=t_{N-1}$. 

Among time interval $(i, j)$, if the storage with EDCR bid has an extra discharge action at time $a$ before  $j$, \ie $i<a<j$ and $\omega^{d*}_{a}\geq 0, \omega^{c*}_{a} = 0$. From the proof for $t=t_N$ at the beginning, we know that the storage action stays the same at $t=j$ when switching to the EDCR bid. So, with EDCR bid, the storage must have another charge action at $b$ and $a<b<j$. To summarize, the storage with EDCR bid will have an even number of actions with a charge action following a discharge action sequentially. Lemma~\ref{lemma:CoupleCD} shows that those extra actions under EDCR bid have 
\beq
\omega^{c*}_{b}=-(\pi_{b}+\kappa^c)/(\eta\tau) + (\pi_{j} -\kappa^d)/\tau \geq 0.
\eeq
\beq
\omega^{d*}_{a}=-(\pi_{b}+\kappa^c)/(\eta\tau) + (\pi_{a} -\kappa^d)/\tau \geq 0.
\eeq
\beq
\omega^{c*}_{i}=-(\pi_{i}+\kappa^c)/(\eta\tau) + (\pi_{a} -\kappa^d)/\tau \geq 0.
\eeq
So considering extra actions at time $a$ and $b$, we can get the KKT condition of the dispatch action at time $t_{N-1}$ for EDCR bid with the equations above. And the optimal storage dispatch under EDCR bid also charges $\overline{E}$ at time  $t_{N-1}$. The same argument can be applied to the case with more extra actions under the EDCR bid.

For case \underline{2) discharge to $\underline{E}$ at $t_{N-1}$ and charge to $\overline{E}$ at $t_{N}$}, the same proof can be applied.

\underline{Among time interval $[t_{i-1}, t_i]$, $\forall i \in \{2,...,N\}$}, we can use the proof of  time interval $[t_{N-1}, t_N]$ to gradually prove from $i=N$ to $i=2$. Then we have the storage with EDCR bid has the same actions as that with SoC independent bid over the time intervals in set ${\cal A}$.

Until now, we've proved that the storage has the same actions for time intervals in ${\cal A}$. So \underline{only considering intervals in ${\cal A}$}, we have $\phibf(\thetabf^*) = \phibf(\overline{\thetabf})$. The payment that the storage receives and the true storage profit in ${\cal A}$ under the EDCR bid is the same as those under the SoC-independent bid, \ie ${\cal P}(\phibf(\thetabf^*))={\cal P}(\phibf(\overline{\thetabf}))$ and $\Pi(\thetabf^*)=\Pi(\overline{\thetabf})$. From \eqref{cons:c1}, we know that the bid-in cost under the EDCR bid is no higher than that of the SoC-independent bid. Consequently, the bid-in profit under the EDCR bid is no less than that under the latter, \ie $\tilde{\Pi}(\thetabf^*) \geq \tilde{\Pi}(\overline{\thetabf})$.

Next, we prove for \underline{intervals in $[T]\setminus {\cal A}$}, where the bid-in profit and true profit are zeros for storage with SoC-independent bid. Following the individual rationality of convex market clearing\cite[Definintion 1]{Cong&Tong:22Allerton}, if individual units do individual profit maximization given the market clearing price from \eqref{eq:CVX_bicomplete}, they will find the dispatch signal from \eqref{eq:CVX_bicomplete} match their individual  optimal dispatch. Therefore, storage receives the maximum bid-in profit with the dispatch signal from \eqref{eq:CVX_bicomplete}. Known that one feasible solution  for storage is not to have extra actions in $[T]\setminus {\cal A}$. So if the optimal scheduling of storage with EDCR bid has extra optimal dispatch actions in $[T]\setminus {\cal A}$ and there's no dual degeneracy, the total bid-in profit of storage should be positive in $[T]\setminus {\cal A}$. 

Therefore, during $[T]$, if we have $\phibf(\thetabf^*) \neq \phibf(\overline{\thetabf})$, then there exist at least one interval in $[T]\setminus {\cal A}$ such that storage has charging or discharging actions under the EDCR bid. If  there's no dual degeneracy, then the bid-in profit of storage over $[T]$, defined in \eqref{EDCRProf}, has 
\beq
\tilde{\Pi}(\thetabf^*) > \tilde{\Pi}(\overline{\thetabf}).
\eeq

From  Lemma~\ref{eq:SoCcostG}, the true storage cost under the EDCR bid is never underestimated by the bid-in SoC-dependent cost.  So, from \eqref{eq:true cost}, we know that the true storage profit under EDCR bid $\Pi(\thetabf^*)$ over $[T]\setminus {\cal A}$ is nonnegative. Therefore, if we have $\phibf(\thetabf^*) \neq \phibf(\overline{\thetabf})$ over $[T]$ and there's no dual degeneracy, we have
\beq\label{eq:TrueProfitability} 
\Pi(\thetabf^*) > \Pi(\overline{\thetabf}).
\eeq
\qed

\begin{lemma} \label{eq:SoCcostG}
Under A1 and  constraints \eqref{cons:c2},  the true SoC-dependent cost is never underestimated by the storage cost  under the EDCR bid.
\end{lemma}
Proof: Consider the optimal storage schedule under EDCR bid $\ebf^*:=(e^*_t)$, which has $e^*_1=s$ for the initial SoC. When  $\ebf^*$ satisfies A1, we have
\beq
\begin{cases}
e^*_t =s,&  t \in \{ 1,..., \chi\}\\
e^*_t \in \{\underline{E}, \overline{E}\}, &  t \in \{ \chi+1,..., T\}\\
e^*_t \in \Phi, & t=T+1
\end{cases}
\eeq
Therefore, when $t \in \{\chi,..., T\}$, we have $e^*_t\in \{s, \underline{E}, \overline{E}\}$ and $e^*_{t+1} = e^*_t + (q^{c*}_t \eta -  q^{d*}_t)\tau \in \Phi$. With the definition for the set  ${\cal S}:=\{ \tilde{\sbf} := (\tilde{q}^c, \tilde{q}^d,\tilde{e}) | \tilde{e} \in \{s, \underline{E}, \overline{E}\}, \tilde{e} + (\tilde{q}^c \eta -  \tilde{q}^d)\tau \in \Phi\}$, we consider all possible single stage action for storage.  Here we use the same formulation for the single stage storage cost  in \eqref{eq:CVX_single}, but we denote bidding parameters   $\thetabf^*$ as decision variables, and the storage actions are given parameters in set ${\cal S}$. From constraints \eqref{cons:c2}, we know $\tilde{f}(\thetabf^*; \tilde{\sbf}) \geq \psi( \tilde{\sbf}), \forall \tilde{\sbf} \in {\cal S}$.
So, under the EDCR bid, the bid-in storage cost is no less than the true SoC-dependent cost.\qed

\begin{lemma}\label{eq:SoCLemma}
    From the convexified electricity  market clearing problem (\ref{eq:CVX_bicomplete}) together with the dual variables defined, we have
\beq\label{eq:phi-omega}
\varrho^*_t=\sum_{n=t+1}^T(\omega^{d*}_n-\omega^{c*}_n), \forall t \in [T-1], ~~\varrho^*_T = 0.
\eeq
\end{lemma}
Proof: The proof directly follows the KKT condition of (\ref{eq:CVX_bicomplete}). We know that the Lagrangian function taking derivative over SoC $e_{T+1},...,e_2$ gives
\beq
\begin{array}{l}
    \varrho^*_{T} = 0,\\
    -\varrho^*_{T}+\varrho^*_{T-1} - \omega^{d*}_{T} + \omega^{c*}_{T} =0,\\\nn
    ...,\\
    -\varrho^*_{2} + \varrho^*_{1} -  \omega^{d*}_{2} + \omega^{c*}_{2} =0.  
\end{array}
\eeq
From the first equation, $\varrho^*_T = 0$. Summing up the first two equations, $\varrho^*_{T-1} =  \omega^{d*}_{T} -  \omega^{c*}_{T}$. Similarly, by summing up the first equation to the equation containing $\varrho^*_{t}$, we have \eqref{eq:phi-omega}.\qed



\begin{lemma} \label{lemma:CoupleCD}
 Consider optimal scheduling from (\ref{eq:CVX_bi}) for storage with two consecutive actions at time $i$ and $j$ respectively. If storage charges at time $i$ and discharges at $j$, LMP for these two intervals has 
 \beq
 \pi_{i} +\kappa^c-\eta (\pi_{j} -\kappa^d) \le 0. \eeq 
 If the storage  discharges at time $i$ and charges at $j$,   LMP for these two intervals has 
  \beq
  \pi_{j} +\kappa^c-\eta(\pi_{i} -\kappa^d) \le 0. \eeq 
\end{lemma}
Proof: If storage charges at time $i$ and discharges at $j$, the KKT conditions \eqref{eq:EDCRKKTEE}   give 
\beq\label{eq:KKTeq2}
\begin{array}{l}
     \kappa^c +\pi_{i} +(-\sum_{n=i+1}^T(\omega^{d*}_n- \omega_n^{c*})+\omega^{c*}_{i})\eta \tau +\bar{\rho}^{c*}_{i}=0, \\
\kappa^d-\pi_{j} +(\sum_{n=j+1}^T( \omega^{d*}_n- \omega_n^{c*})+\omega^{d*}_{j}) \tau +\bar{\rho}^{d*}_{j}=0. 
 \end{array}
\eeq
 From the complementary slackness condition,   $\underline{\rho}^{c*}_{i}=0$, $\underline{\rho}^{d*}_{j}=0$, $\omega^{c*}_{n}\geq 0$, $\forall n \in \{i,...,j-1\}$, and $\omega^{d*}_{n} =0$, $\forall n \in \{i+1,...,j-1\}$. So sum two equations in \eqref{eq:KKTeq2} up, we have
\beq
\begin{array}{l}
\pi_{i} +\kappa^c-\eta (\pi_{j} -\kappa^d)\\
\le (\sum_{n=i+1}^{j-1}\omega^{d*}_n- \sum_{n=i}^{j}\omega^{c*}_n)\eta\tau = \sum_{n=i}^{j}( - \omega^{c*}_n)\eta\tau \le 0.\nn
\end{array}
\eeq 

We can prove this in the same way if the storage discharges at time $t$ and charges at $m$. 
\qed

\subsection{Simulation parameters}\label{sec:simparamLoad}

The system consists of 19 conventional generators, with energy bids of [$\mathbf{0}_{1\times 4}$, 8:16.96:246] \$/MWh\footnote{The notation, $a:b:c$, represents a vector beginning with $a$, and increasing with a step size $b$ until it goes beyond $c$.}, regulation up bids of [$\mathbf{0}_{1\times 4}$, 4:8.48:123] \$/MWh, and regulation down bids of [3.5:3.5:67] \$/MWh. Generator capacity limits in the energy market and regulation market are $[200 \times \mathbf{1}_{1\times 5}, 150  \times \mathbf{1}_{1\times 13}, 1000]$ MW and $[10  \times  \mathbf{1}_{1\times 18}, 200]$ MW, respectively. The solar generator bids 0 \$/MWh in both the energy and regulation markets. For simplicity, we simulated a single bus system.
 
\begin{figure}[!htb]
   \centering
   \vspace{-0.1in}
   \includegraphics[scale=0.32]{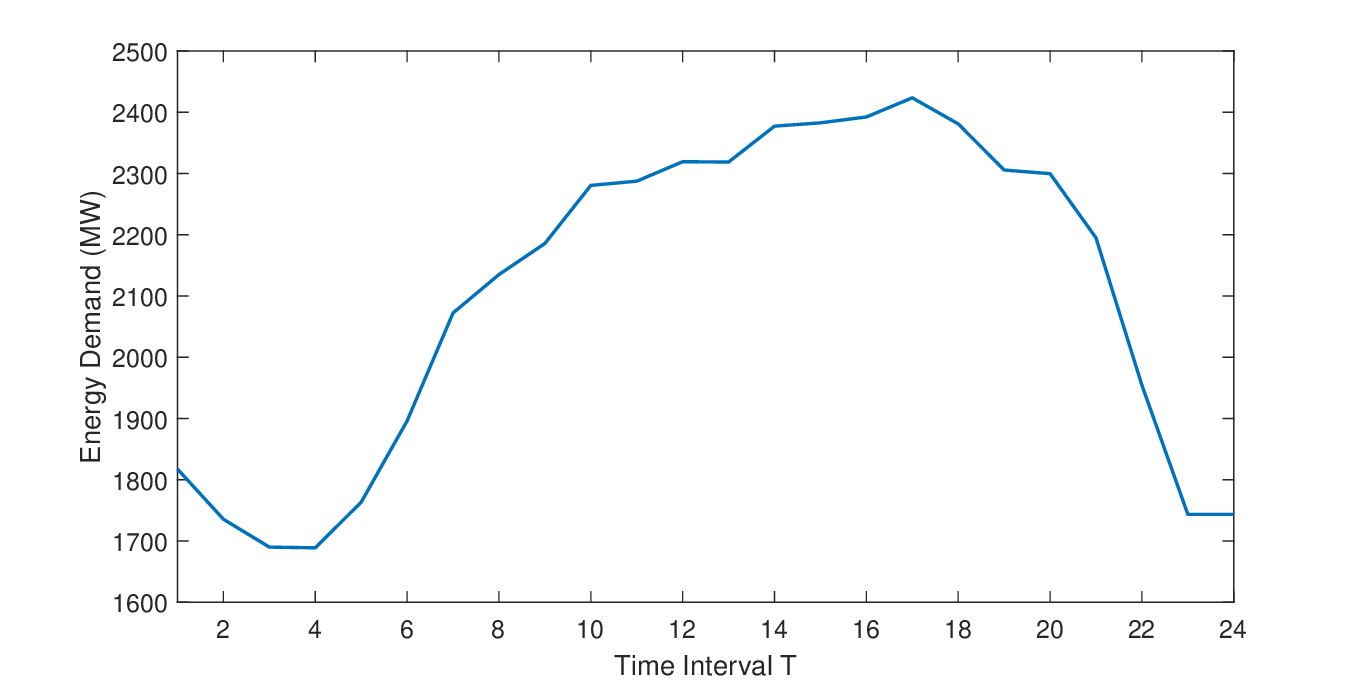} 
\caption{\scriptsize Energy demand in the simulation.}
\label{fig:Energydemand}
\end{figure}

In the energy market, the mean value of load demand $\pmb{d}$ is shown in Fig.~\ref{fig:Energydemand}, which came from scaling the average demand at ISO New England\cite{ISONE:21}. The noise added to the energy demand follows the Gaussian distribution ${\cal N}(0, \frac{d_t}{100})$. In the one-shot dispatch, the simulation period spans from $t=5$ to $t=11$, while the rolling-window simulation covers the entire $24-T$ period.

\begin{figure}[!htb]
   \centering
   \vspace{-0.1in}
   \includegraphics[scale=0.4]{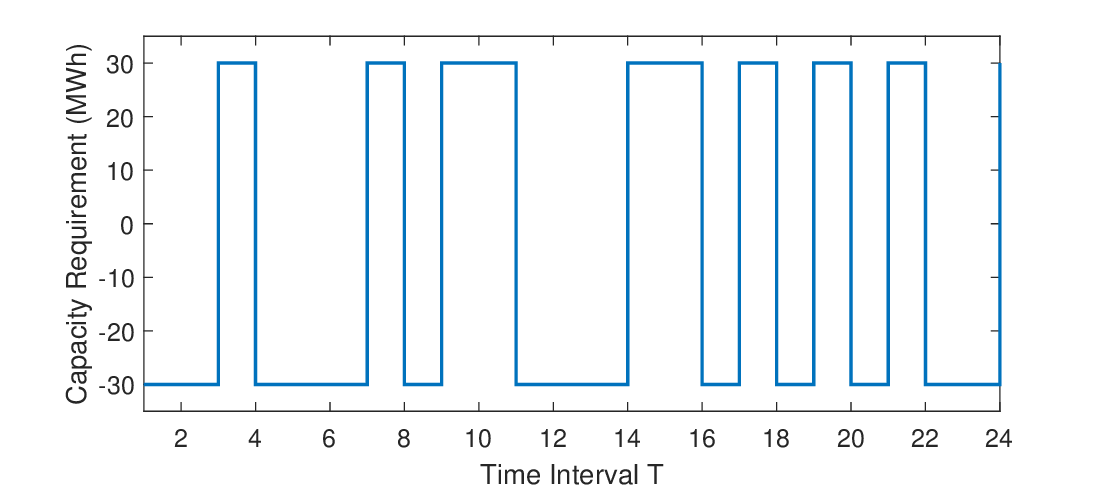} 
\caption{\scriptsize Regulation capacity demand in the simulation. Positive values on the y-axis represent $\xi^u_t$, while negative values equal $-\xi^d_t$ for different intervals.}
\label{fig:RegRequire}
\end{figure}

According to the real-time regulation market results of PJM \cite{PJM:DataMiner}, the required amount of regulation capacities remains stable, while the clear price for regulation capacities exhibits volatility. To simulate such a phenomenon, we set the regulation capacity demand to be constant over time, as shown in Fig.~\ref{fig:RegRequire}. In the figure, positive values represent regulation up requirements, while negative values represent regulation down requirements. Meanwhile, we introduce fluctuations in the maximum output of the solar generator to simulate the volatility of the clearing price for regulation capacity.

\begin{figure}[!htb]
   \centering
   \vspace{-0.1in}
   \includegraphics[scale=0.36]{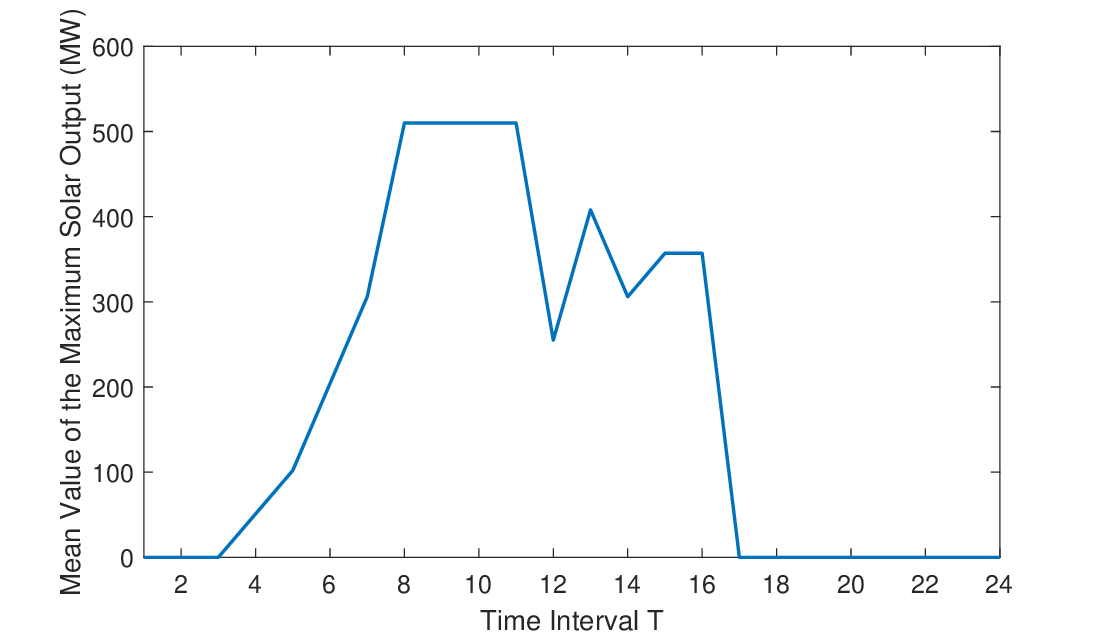}
\caption{\scriptsize Mean Value of the Maximum Solar Output.}
\label{fig:Solar}
\end{figure}

In the simulation, the maximum output of the solar generator varies over time. The mean value of solar output $\pmb{\mu}$ for each interval during the 24-T period is illustrated in Fig.~\ref{fig:Solar}. The volatility of its power output at time $t$ follows ${\cal N}(0, \frac{\mu_t}{1000})$.

%% file: simadd_v1.tex
\subsection{Energy market simulation}\label{sec:EMsim}

We considered 31 generators with Generator bids [10:5:160] \$/MWh and generation capacities of 30 MW.  In this single bus system, we had one storage with the same capacity limit and initial SoC in Sec.\ref{sec:Ex}. We took the bidding parameter for storage  in the top left of Fig.~\ref{fig:ToyEx} and multiplied it by 15 to conduct this simulation. Storage only participated in the energy market in this simulation. We took the 24-hour inelastic demand in Fig.~\ref{fig:Energydemand} and multiplied all demand by 0.1. We adopted the same random sample generation methods in Sec.~\ref{sec:simparamLoad} and simulated 500 scenarios for this section. We scaled the inelastic demand by adding all scenarios with the demand scale factor, and simulation results are shown below in Fig.~\ref{fig:EnergyMarket} with  the x-axis representing the demand scale factor.

 \begin{figure}[!htb]
   \centering
   \vspace{-0.1in}
\includegraphics[scale=0.41]{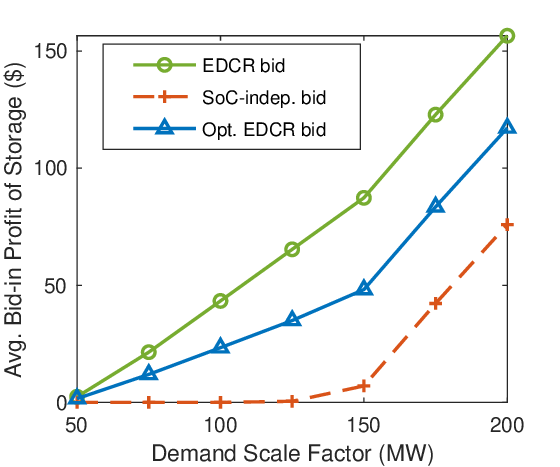}  
\includegraphics[scale=0.41]{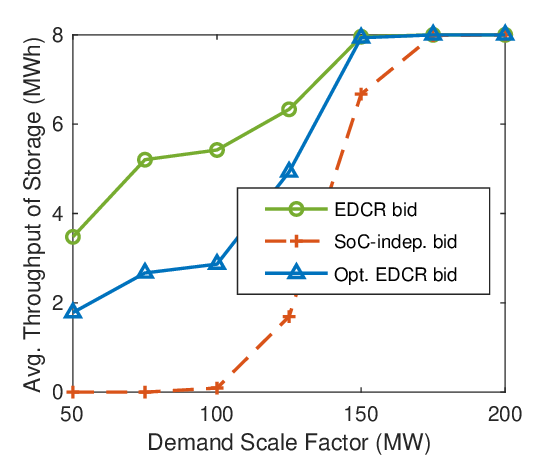}  
\includegraphics[scale=0.41]{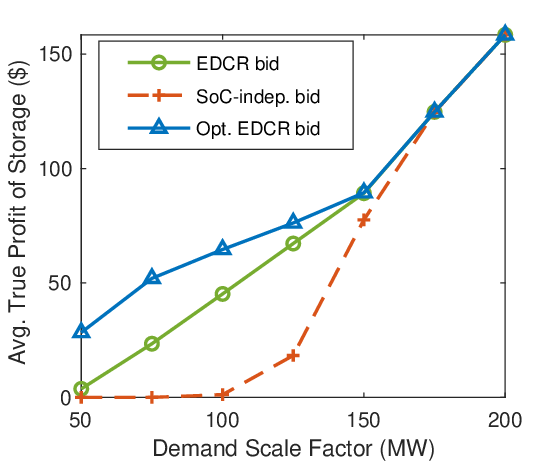}    
\caption{\scriptsize Simulation results. (Top left: storage throughput; top right: storage bid-in profit).}
\label{fig:EnergyMarket}
\end{figure}

We drew the same conclusions in the main text simulation section. The storage utilization, bid-in profit, and true profit were higher when submitting the EDCR bids. Storage submitting the Opt. EDCR bid has a higher true profit than the EDCR bid. All metrics for storage increase when the demand scale factor increases because the  mean and volatility of LMP increase. 